\documentclass{article}

\usepackage{url,hyperref,lineno,microtype,subcaption}
\usepackage{amsmath,amsfonts,amsthm,thmtools,graphicx}

\declaretheorem[style=plain,name=Theorem,numberwithin=section]{theorem}

\begin{document}

\title{Identifying efficient controls of complex interaction networks using genetic algorithms}

\author{
    Victor-Bogdan Popescu
    \thanks{
        Computational Biomodelling Laboratory, Turku Center for Computer Science, Turku, Finland;
        Computer Science Unit, \r{A}bo Akademi University, Turku, Finland
    }
    \and Krishna Kanhaiya
    \thanks{
        Computational Biomodelling Laboratory, Turku Center for Computer Science, Turku, Finland;
        Computer Science Unit, \r{A}bo Akademi University, Turku, Finland;
        Finnish Red Cross Blood Services, Helsinki, Finland
    }
    \and Iulian N\u{a}stac
    \thanks{
        Faculty of Electronics, Telecommunications and Information Technology, University Politehnica of Bucharest, Romania
    }
    \and Eugen Czeizler
    \thanks{
        Computational Biomodelling Laboratory, Turku Center for Computer Science, Turku, Finland;
        Computer Science Unit, \r{A}bo Akademi University, Turku, Finland;
        National Institute for Research and Development in Biological Sciences, Bucharest, Romania
    }
    \and Ion Petre
    \thanks{
        Computational Biomodelling Laboratory, Turku Center for Computer Science, Turku, Finland;
        Department of Mathematics and Statistics, University of Turku, Finland;
        National Institute for Research and Development in Biological Sciences, Bucharest, Romania;
        \textit{ion.petre@utu.fi}
    }
}

\date{July 2020}

\maketitle

\begin{abstract}

    Control theory has seen recently impactful applications in network science, especially in connections with applications in network medicine. A key topic of research is that of finding minimal external interventions that offer control over the dynamics of a given network, a problem known as  network controllability. We propose in this article a new solution for this problem based on genetic algorithms. We tailor our solution for applications in computational drug repurposing, seeking to maximise its use of FDA-approved drug targets in a given disease-specific protein-protein interaction network. We show how our algorithm identifies a number of potentially efficient drugs for breast, ovarian, and pancreatic cancer. We demonstrate our algorithm on several benchmark networks from cancer medicine, social networks, electronic circuits, and several random networks with their edges distributed according to the Erd\H{o}s-R\'{e}nyi, the small-world, and the scale-free properties. Overall, we show that our new algorithm is more efficient in identifying relevant drug targets in a disease network, advancing the computational solutions needed for new therapeutic and drug repurposing approaches.

    \textbf{Keywords}: network medicine, drug repurposing, network controllability, cancer medicine, genetic algorithm

\end{abstract}

\section{Introduction}

Network modelling in systems medicine has emerged as a powerful analytics approach in the last couple of decades (\cite{Kitano:2002aa, Saqi:2016aa, Tian:2012aa}). Its aim is to analyse diseases and drug interventions as ways of acting, and re-acting, over bio-medical dynamical networks (\cite{Barabasi:2011aa, Goh:2007aa, Zhou:2014aa}), such as the protein-protein interaction networks (\cite{art-cgkkp18, art-kcgp17}), signalling networks (\cite{Ochsner:2019aa}), metabolic networks (\cite{Misselbeck:2019aa}), and immunological responses (\cite{Davis:2017aa}). In this framework, a disease is seen as emerging from some of its modules being affected (directly or through cascading signals) and from critical nodes in the network being deregulated (\cite{Liu:2019aa}). Similarly, drug therapies are seen as outside  controlled interventions within a deregulated network with the aim of either re-balancing the system or possibly isolating some specific components of the network (\cite{Cheng:2019aa}). A particular advantage of this approach is reasoning about multiple-drug interventions, analysing and predicting multi-drug synergies, as well as aiming for personalised therapies. The current ``state'' of a patient can be reflected in its personalised network (\cite{Tian:2012aa}), by integrating elements specific to the disease, to treatment pathways, and to the patient herself, such as genetic mutations and current medical conditions and treatments.

Instead of acting over each individual disregulated component, one can try to influence several of these entities through a few well-chosen interventions, and to have them spread in cascade into the network using the network's own internal interconnections. It turns out that network controllability is a topic of high relevance in this area with a rich theory to support it (\cite{Kalman1963}). It has found in recent years powerful applications in computational systems medicine and therapeutics (\cite{art-lsb11}, \cite{art-kcgp17, art-cgkkp18, Zhang:2017aa, Liu:2015aa, Gao:2014aa, Guo:2017aa}).

The theory of network controllability aims at providing sound and theoretically accurate description of what control means within a network, and how it can be achieved. Intuitively, achieving control over a system from a set of input nodes means being able to drive that system from any initial setup to any desired state. This is an intrinsic optimisation problem with the objective to minimise the number of input nodes (e.g., drug targets) needed for the control. Additional constraints may be added depending on the application, such as the control pathways from the input nodes to the controlled nodes to be short, or the input nodes to be primarily selected from a given set of preferred nodes (targets of standard therapy drugs). This leads to several problem variations, such as: structural controllability (\cite{art-lsb11}), i.e., identifying  pathways that offer control over the system regardless of its numerical setup; target controllability (\cite{Gao:2014aa}), i.e., achieving control over a predefined set of target nodes; minimum dominating sets (\cite{Molnar:2013aa}), i.e., finding a minimal set of nodes that are one step upstream of all other nodes in the network. Some of these optimisation problems are known to have efficient algorithmic solutions (\cite{art-lsb11}). Others, on the contrary, are known to be computationally difficult, yet approximate efficient solutions are still achievable (\cite{art-cgkkp18}). 

Motivated by the applicability of network control in systems medicine, the problem we focus on in this paper is minimising the number of external interventions needed to achieve target control of a system. We are particularly interested in the case where the targets are disease-specific survivability-essential genes, key targets for synthetic lethality (\cite{Rancati:2018aa}). We identify control interventions that are achievable through the delivery of FDA-approved drugs, by giving a preference to FDA-approved drug targets being selected as input nodes. The target controllability problem is known to be NP-hard, meaning that finding the smallest set of inputs for controlling the target set is computationally prohibitive for large networks. We give as a solution an  approximation of the minimal solution based on genetic algorithms,  well known heuristic choices for nonlinear optimisation problems (\cite{Whitley:2012aa}). We demonstrate that this approach offers an efficient solution for applications in combinatorial drug therapy identification and drug repurposing.

\section{Materials and methods}

\subsection{Network controllability}

We introduce briefly the basic concepts of network controllability and the Kalman condition for the target controllability problem. For more details we refer to \cite{art-cgkkp18}. By convention, all vectors are considered to be column vectors so that the matrix-vector multiplications are well defined.

Let $A\in \mathbb{R}^{n\times n}$ be an $n\times n$ matrix, for some $n>1$. The linear dynamical system defined by matrix $A$ is an $n$-dimensional vector $x$ of real functions, $x:\mathbb{R}\rightarrow \mathbb{R}^n$, defined as the solution of the system of ordinary differential equations

\begin{equation}
    \frac{dx(t)}{dt}=Ax(t),
\end{equation}

for some given $x(0)$. The structure of a linear dynamical system can be thought of as an edge-labeled directed graph with $n$ vertices and adjacency matrix $A$: for any nodes $i,j$, $1\leq i,j\leq n$, the (directed) edge $(i,j)$ documents node $i$ being influenced by node $j$, with the weight of the influence (documented as the label of edge $(i,j)$) given by the $(i,j)$ entry $a_{i,j}$ of matrix $A$. 

We consider a subset of input nodes $I\subseteq \{1,2,\ldots, n\}$, $I=\{i_1,\ldots,i_m\}$, $1\leq m\leq n$, thought of as the nodes of the linear dynamical system on which an external contribution can be applied to influence the dynamics of the system. The subset of input nodes $I$ can also be described through its characteristic matrix $B_I\in\mathbb{R}^{n\times m}$, defined as follows: $B_I(r,s)=1$ if $r=i_s$ and $B_I(r,s)=0$ otherwise, for all $1\leq r\leq n$ and $1\leq s\leq m$. The external influence is exerted through an $m$-dimensional input vector $u$ of real functions, $u:\mathbb{R}\rightarrow\mathbb{R}^m$. The influence of the input vector $u$ on the linear dynamical system is described by the equation

\begin{equation}
    \label{equation-lds-input}
    \frac{dx(t)}{dt}=Ax(t)+B_Iu(t).
\end{equation}

We also consider a subset of target nodes $T\subseteq\{1,2,\ldots,n\}$, $T=\{t_1,\ldots,t_l\}$, $m\leq l\leq n$, thought of as a subset of the nodes of the linear dynamical system whose dynamics we aim to control (as defined below) through a suitable choice of input nodes and of an input vector. The subset of target nodes can also be defined through its characteristic matrix $C_T\in \mathbb{R}^{l\times n}$, defined as follows: $C_T(r,s)=1$ if $t_r=s$ and $B_I(r,s)=0$ otherwise, for all $1\leq r\leq l$ and $1\leq s\leq n$.

The triplet $(A,I,T)$ is called the targeted linear dynamical system with inputs, defined by matrix $A$, input set $I$ and target set $T$. We say that this system is target controllable if for any $x(0)\in\mathbb{R}^n$ and any $\alpha\in\mathbb{R}^l$, there is an input vector $u:\mathbb{R}\rightarrow\mathbb{R}^m$ such that the solution $\tilde{x}$ of \eqref{equation-lds-input} eventually coincides with $\alpha$ on its $T$-components, i.e., $C_T\tilde{x}(\tau)=\alpha$, for some $\tau\geq 0$. Intuitively, the system being target controllable means that for any input state $x_0$ and any desired final state $\alpha$ of the target nodes, there is a suitable input vector $u$ driving the target nodes to $\alpha$. Obviously, the input vector $u$ depends on $x_0$ and $\alpha$. 

\begin{figure}[htb]
    \begin{center}
        \includegraphics[width=0.9\textwidth]{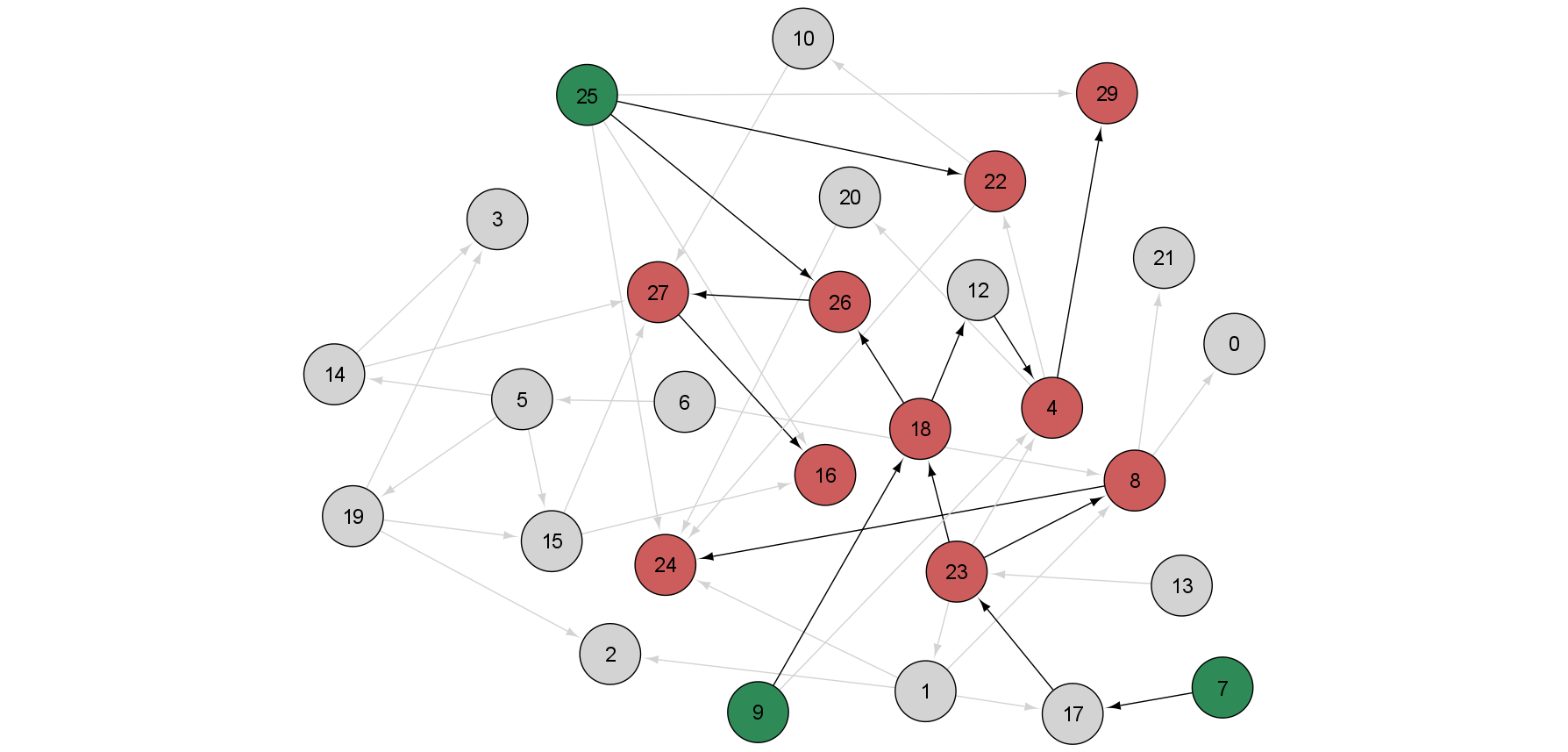}
    \end{center}
    \caption{The structural setup of the target controllability problem. In green: input nodes. In red: target nodes. The control paths are indicated with thicker arrows.}
    \label{figure-structural-setup}
\end{figure}

We illustrate in Figure \ref{figure-structural-setup} the structural setup of the target controllability problem.

The question whether a given targeted linear dynamical system with inputs $(A,I,T)$ is controllable has an elegant algebraic answer known as the Kalman condition.

\begin{theorem}[\cite{Kalman1963}]
	A targeted linear dynamical system with inputs $(A,I,T)$ is controllable if and only if its controllability matrix $[C_TB_I,C_TAB_I,C_TA^2B_I,\ldots,$ $C_TA^{n-1}B_I]$ is of full rank.
\end{theorem}

The controllability matrix of the targeted linear dynamical system with inputs $(A,I,T)$ is an $l\times mn$ matrix, meaning that being of full rank is equivalent with its rank being equal to $l$ (since $l\leq mn$). Intuitively, this matrix describes all weighted paths from the input nodes to the target nodes in directed graph associated to the linear dynamical system described by matrix $A$. This line of thought can be further developed into a structural formulation of the targeted controllability and into a graph-based solution for it, see \cite{art-cgkkp18}.

The problem we focus on and solve in this paper is that of minimising the set of input nodes needed for the target controllability of a dynamical system. For the linear dynamical system defined by a matrix $A\in\mathbb{R}^{n\times n}$ and a set of target nodes $T$ of size $l$, the problem is to find the smallest $m$, $1\leq m\leq l$, such that for a suitable input set $I$ of size $m$, the targeted linear dynamical system with inputs $(A,I,T)$ is target controllable.

We add an extra layer of optimisation to the target controllability problem, motivated by medicine as our application domain. In medical applications, the input functions mimic the effect of drug delivery, with the input nodes being targets of commercially available drugs. Consequently, we introduce in our mathematical formulation an additional set of so-called preferred nodes $P\subseteq \{1,2,\ldots, n\}$, with the aim of selecting in the input set $I$ as many preferred nodes as possible. The problem in this case becomes the following. For the linear dynamical system defined by a matrix $A\in\mathbb{R}^{n\times n}$, a set of target nodes $T$ and a set of preferred nodes $P$, the problem is to find a smallest-sized input set $I$ whose intersection with $P$ is maximal, such that the targeted linear dynamical system with inputs $(A,I,T)$ is controllable, i.e., such that matrix $[C_TB_I,C_TAB_I,C_TA^2B_I,\ldots,$ $C_TA^{n-1}B_I]$ is of full rank.

The optimisation version of the above-defined target controllability problem has been shown to be NP-hard in \cite{art-kcgp17}, meaning that an exact solution may require a prohibitive amount of time, exponential in the number of nodes. We focus in this paper on the next most-feasible objective, an efficient and effective heuristic solution that comes, however, with no guarantee of being optimal.

\subsection{The outline of the genetic algorithm}

The algorithm takes as input a network given as a directed graph $G=(V,E)$ and a list of target nodes $T\subseteq V$, $T=\{t_1,\ldots,t_l\}$. We denote the graph's adjacency matrix by $A_G$. The algorithm gives as a result a set of input nodes $I\subseteq V$ controlling the set $T$, with the objective being to minimise the size of $I$. The algorithm can also take as an additional, optional input a set $P\subseteq V$ of so-called preferred nodes. In this case, the algorithm will aim for a double optimisation objective: minimise the set $I$, while maximising the number of elements from $P$ included in $I$. Our typical application scenario will be that of a network $G$ consisting of protein-protein interactions specific to a disease mechanism of interest, with the set of targets $T$ being a disease-specific set of essential genes, and the set of preferred nodes $P$ a set of proteins targetable by available drugs or by specially designed compounds (e.g., inhibitors, small silencing molecules, etc.) The terminology we use to describe the algorithm, e.g., population/chromosome/crossover/mutation/fitness is standard in the genetic algorithm literature and refers to its conventions, rather than being suggestive of specifics in molecular biology. 

Our algorithm will start by generating several solutions to the control problem, in the form of several control sets $I_1, \ldots, I_m$ -- we discuss in Section \ref{subsection-chromosome} how this is achieved. Each such solution is encoded as a ``chromosome'', i.e., as a vector of (not necessarily distinct) ``genes'' $[g_1,\ldots,g_l]$, where for all $1\leq i\leq l$, $g_i\in V$ controls the target node $t_i\in T$. In particular, $g_i$ is an ancestor of $t_i$ in graph $G$, for all $1\leq i\leq l$.

A set of chromosomes is called a population. Note that a chromosome will always encode a solution to our optimisation problem, throughout the iterative run of the algorithm. Any population maintained by the algorithm consists of several such chromosomes, some better than others from the point of view of our optimisation criteria, but all valid solution to the target controllability problem to be solved. 

The algorithm iteratively generates successive populations (sets of chromosomes) that get better at the optimisation it aims to solve: the size of the control set gets smaller and the proportion of preferred nodes in the control set gets higher. The algorithm stops after a maximum number of iterations, or after a number of iterations in which the quality of the solution does not get improved. This pre-defined stop is necessary since the target controllability problem is known to be NP-hard and so, finding the optimal solution can require a prohibitively high number of steps, potentially exponential in the number of nodes in the network. The end result consists of several solutions to the problem, represented by all the control sets in the final population obtained by the algorithm. 

The initial population of solutions is randomly generated in such a way that each element selected for it is indeed a solution to the target controllability problem $(A_G,I,T)$. To generate the next generation/population from the current one, we use three techniques.

\begin{itemize}
    \item Retain in the population the best solutions (from the point of view of the optimisation problem to be solved). ``Elitism'' will be used to conserve the best solutions (discussed in the next sections in details).
    \item Add random chromosomes (all being valid solutions to the optimisation problem, albeit potentially of lower fitness score than some of the others in the population).
    \item Generate new solutions/chromosomes resulting from combinations of those in the current population. A selection operator is used to choose the chromosomes which will produce offsprings for the following generation. New chromosomes are produced using crossover and mutation (discussed in the next sections in details).
\end{itemize}

A list of all the parameters used by the genetic algorithm can be found in Table \ref{table-parameters}. The basic outline of the proposed genetic algorithm is described below. All operators will be detailed in the following subsections.

\begin{table}[htb]
    \begin{center}
        \begin{tabular}{|p{.15\textwidth}|p{.40\textwidth}|p{.15\textwidth}|p{.15\textwidth}|}
            \hline
            \textbf{Parameter} & \textbf{Meaning} & \textbf{Type} & \textbf{Default value} \\
            \hline
            $N$ & \raggedright{Total number of generations for which the algorithm will run} & $N \in \mathbb{N}_{\geq 1}$ & $N = 10000$ \\
            \hline
            $n$ & \raggedright{Total number of chromosomes in a generation} & $n \in \mathbb{N}_{\geq 2}$ & $n = 80$\\
            \hline
            $p_m$ & \raggedright{Probability of mutation for a chromosome} & $p_e \in \mathbb{R}_{[0, 1]}$ & $p_m = 0.25$ \\
            \hline
            $p_e$ & \raggedright{Maximum percentage of elites in a generation} & $p_r \in \mathbb{R}_{[0, 1]}$ & $p_e = 0.01$ \\
            \hline
            $p_r$ & \raggedright{Percentage of randomly generated chromosomes in a generation} & $p_m \in \mathbb{R}_{[0, 1]}$ & $p_r = 0.25$ \\
            \hline
            $\text{max}_{path}$ & \raggedright{Maximum number of interactions in a control path} & $\text{max}_{path} \in \mathbb{N}_{\geq 1}$ & $\text{max}_{path} = 5$ \\
            \hline
            $\text{max}_{rand}$ & \raggedright{Maximum number of randomly generated genes in a chromosome} & $\text{max}_{rand} \in \mathbb{N}_{\geq 1}$ & $\text{max}_{rand} = 15$ \\
            \hline
        \end{tabular}
    \end{center}
    \caption{The parameters used by the genetic algorithm.  $\textbf{N}$ represents the maximum number of generations for which the algorithm will run. Additionally, the algorithm will stop after $1000$ generations with no improvement in the fitness of the best chromosomes. Higher values for $N$ improve the chances of getting better results, but increase the overall running time. $\textbf{n}$ represents the total number of chromosomes in a generation. This includes randomly generated chromosomes, elite chromosomes of the previous generation, as well as offspring of the chromosomes in the previous generation. Higher values result in a larger population, thus increasing the gene pool, however it might spread the search space too much, getting the process closer to a random search. $\textbf{p}_m$ represents the probability of mutation for a chromosome. It is defined for an entire chromosome, and not for a particular gene. Increasing its value will help with the exploration of the solution space, but too much will make the process get closer to a random search. A value of $0$ will deactivate the mutation operator. $\textbf{p}_e$ represents the percentage of elite chromosomes in a population. Higher values will increase the number of preserved chromosomes with high fitness score over the generations, but the solution space will get smaller. A value of $0$ will deactivate elitism. $\textbf{p}_r$ represents the percentage of new randomly generated chromosomes in a population. Higher values will increase the exploration of the solution space, but it will also be getting the process closer to a random search. A value of $0$ will deactivate this feature.}
    \label{table-parameters}
\end{table}

\begin{enumerate}
    \item \emph{Generate the initial population.} We set $t \leftarrow 0$ for the first generation. We initialise $P(t)$ with a number of $n$ randomly generated chromosomes.
    \item \emph{Preserve the fittest chromosomes.} \label{algorithm-outline-step-2} We evaluate the fitness of all chromosomes in $P(t)$. We add to the next population $P_{t+1}$ the $p_e \cdot n$  chromosomes in the current generation with the highest fitness score, where $0\leq p_e<1$ is the `elitism' parameter. If there are more chromosomes of equal fitness being considered, the ones to be added are randomly chosen.
    \item \emph{Add random chromosomes.} We add  $p_r \cdot n$ new randomly generated chromosomes to $P(t + 1)$, where $0\leq p_r<1$ is the `randomness' parameter.
    \item \emph{Add the offsprings of the current population.} We apply two times the selection operator on $P(t)$, obtaining two chromosomes of $P_t$ selected randomly with a probability proportional to their fitness score. On the two thus selected chromosomes we apply the crossover operator, obtaining an offspring to be added to $P(t + 1)$. The offspring is added in a mutated form with the mutation probability $0\leq p_m<1$. We continue applying this step until the number of chromosomes in $P(t + 1)$ becomes $n$.
    \item \emph{Iterate.} If the current index $t$ is smaller than the maximum number of generations $N$, then we set $t \leftarrow t + 1$ and we continue with Step \ref{algorithm-outline-step-2}.
    \item \emph{Output.} We return the fittest chromosomes in the current generation as solutions to the problem and we stop the algorithm.
\end{enumerate}

\subsection{Chromosome encoding and the fitness function}
\label{subsection-chromosome}

A chromosome $I$ consists of a vector of $l$ genes $I=[g_1,\ldots, g_l]$, not necessarily of distinct value, where $l$ is the size of the target set $T$. As discussed before, for all $1\leq i\leq l$, $g_i\in V$ controls node $t_i\in T$ and so, in particular, it is an ancestor of $t_i$ in graph $G$. Keeping with our focus on applications in medicine, where paths encode signalling networks, we aim for short paths between the input nodes $g_i$ and the target nodes $t_i$. The maximum allowed length for such a path is encoded in the parameter $\text{max}_{path}$. Any time we discuss an ancestor $g$ of node $t$ we mean implicitly that the shortest path from $g$ to $t$ is of length at most $\text{max}_{path}$.

A chromosome $I=[g_1,\ldots, g_l]$ is always checked to ensure that the nodes encoded by its genes are able to control the target set $T$. This is equivalent to the Kalman matrix corresponding to graph $G$, input set $I$ and target set $T$ having maximum rank $l$. All populations, throughout all steps of the algorithm, will consist only of such chromosomes.

The fitness score of a chromosome $I$ is defined as the complement of the number of distinct nodes encoded by its genes:

\begin{equation}
    f(I) = \frac{(l + 1) - |\text{supp}\{b_i | b_i \in I\}|}{l} \cdot 100.
\end{equation} 

Thus, $0 < f(I) \leq 100$. Considering that we are interested in the smallest possible number of input nodes, the higher the fitness score of a chromosome, the better its encoded solution is.

To generate a random chromosome, we will first initialise each of its $l$ elements $g_i$, $1\leq i\leq l$, with its corresponding target node $t_i$. Then, for a number of randomly selected genes (as many as indicated by the parameter $\text{max}_{rand}$), we will replace gene $g_i$ with a randomly chosen ancestor of $t_i$ (at distance at most $\text{max}_{path}$ from $t_i$). Each vector of nodes generated in this way is checked for its Kalman condition: if satisfied, the vector encodes a set of nodes controlling the target set $T$ and it is accepted as a valid output.

\subsection{Selection}

The selection operator is used to choose which chromosomes in the current generation will contribute offsprings to the next generation. The selection of a chromosome depends only on its fitness in relation to the average fitness of the current generation: the better the fitness, the higher the chance it has to be selected. If the current population consists of chromosomes $I_1,\ldots,I_n$, then the probability of selecting chromosome $I_i$, $1\leq i\leq n$ is

\begin{equation}
    p(I_i) = \frac{f(I_i)}{\sum_{k = 1}^{n} f(I_k)}.
\end{equation}

Obviously, $0 < p(I_i)<1$ for any $1\leq i \leq n$, so all chromosomes have a chance of being selected.

\subsection{Crossover}

The crossover operator is used to produce a new (valid) ``offspring'' chromosome from two ``parent'' chromosomes. Each of the offspring chromosome's genes will be directly inherited from one of the two parent chromosomes. The actual parent who will contribute a certain gene is randomly chosen based on the number of occurrences of that gene in the two parent chromosomes: the more often it occurs (in other words, the more efficient its control over the target set), the higher the probability it will be selected for the offspring. For a chromosome $I=[g_1,\ldots,g_l]$ and a gene $g$, we define the number of occurences of $g$ in $I$ to be $\#_I(g)=|\{1\leq i\leq l\mid g_i=g\}|$. Also, for two chromosomes $I,I'$, the number of occurrences of $g$ in $I,I'$ is defined to be $\#_{I,I'}(g)=\#_I(g)+\#_{I'}(g)$.

Consider the two parent chromosomes to be $I_1 = [g_{11}, g_{12}, \ldots g_{1l}]$ and $I_2 = [g_{21}, g_{22}, \ldots g_{2l}]$. For all $1\leq i\leq l$, gene $g_i$ of the offspring will be either $g_{1i}$ or $g_{2i}$. If the genes $g_{1i}$ or $g_{2i}$ are either both preferred (i.e., elements of $P$), or they are both un-preferred, then the probability distribution is defined to be

\begin{equation}
  p(g_{1i}) = \frac{\#_{I_1,I_2}(g_{1i})}{\#_{I_1,I_2}(g_{1i})+\#_{I_1,I_2}(g_{2i})}, \ 
  p(g_{2i}) = \frac{\#_{I_1,I_2}(g_{2i})}{\#_{I_1,I_2}(g_{1i})+\#_{I_1,I_2}(g_{2i})}.
\end{equation}

On the other hand, if one of the two parent genes, say $g_{1i}$, is a preferred node, while the other is not, we reflect our preference for nodes in $P$ in the selection probability in the following way

\begin{equation}
  p(g_{1i}) = \frac{2*\#_{I_1,I_2}(g_{1i})}{2*\#_{I_1,I_2}(g_{1i})+\#_{I_1,I_2}(g_{2i})}, \ 
  p(g_{2i}) = \frac{\#_{I_1,I_2}(g_{2i})}{2*\#_{I_1,I_2}(g_{1i})+\#_{I_1,I_2}(g_{2i})}.
\end{equation}

If the set of nodes obtained as a result of the selection above does not satisfy the Kalman condition, then we do not accept it as a valid solution and we discard it, restarting the crossover operator by selecting two new parent chromosomes. In our numerical experiments, relatively few sets of nodes thus selected failed to satisfy the Kalman condition and this step did not become a bottleneck in our algorithm. 

\subsection{Mutation}

The mutation operator is used to change the values of a small number of genes in a chromosome. The probability for a gene of a chromosome to be selected for mutation is given by the parameter $0\leq p_m<1$. Thus, on average, each newly generated offspring chromosome will have a number of $p_m \times l$ mutated genes. Each gene $g_i$, $1\leq i\leq l$ represents an ancestor of its corresponding target node $t_i$, so  $g_i$ getting mutated corresponds to replacing it with another ancestor of $t_i$; the option of getting $g_i$ again is allowed. The new ancestor is selected randomly from the set of predecessors of $t_i$, with those being preferred nodes having double the weight.

If the newly obtained chromosome is not valid according to the Kalman condition, then we repeat the process with the same genes selected for mutation.

\subsection{Implementation}

The proposed algorithm has been implemented as a cross-platform stand-alone desktop application written in C\# / .NET Core and usable within a command-line interface or under a browser-based graphical interface. The source code and the latest release are available at \cite{sof-p19}.

For each run, the software requires several files -- one containing the list of directed edges (by default, each edge on a separate row, containing semicolon-separated source and target nodes), the list of target nodes (by default, each node on a separate row) and the set of parameters in Table \ref{table-parameters} (by default, as a JSON file). The required format for each of these files is presented in the supplementary information. In addition, a file containing the list of preferred nodes (by default, each node on a separate row) can also be given as an optional input. For the command-line interface, the input data is provided as paths to the corresponding input files. For the graphical user interface, the same options are available, with the added possibility to directly type in or edit the data within the corresponding text fields of the interface. Both cases return the same type of output data, a JSON file containing all of the relevant information of the algorithm run, such as details about the input data, the used parameters, the time elapsed and the control nodes corresponding to each of the identified solutions.

All matrix operations use the Math.NET Numerics library \cite{sof-mn19}. We parallelised the execution of the most used methods, such as chromosome initialisation or crossover. In turn, this required adapting the default pseudo-random number generator into a thread-safe version by using a thread-safe collection of seeds based on the initial random seed.

The graphical interface of the program can be seen in Figure \ref{figure-screen}. Further details on the implementation and usage can be found in the supplementary information and in the GitHub repository (\cite{sof-p19}).

\begin{figure}[htb]
    \begin{center}
        \includegraphics[width=0.9\textwidth]{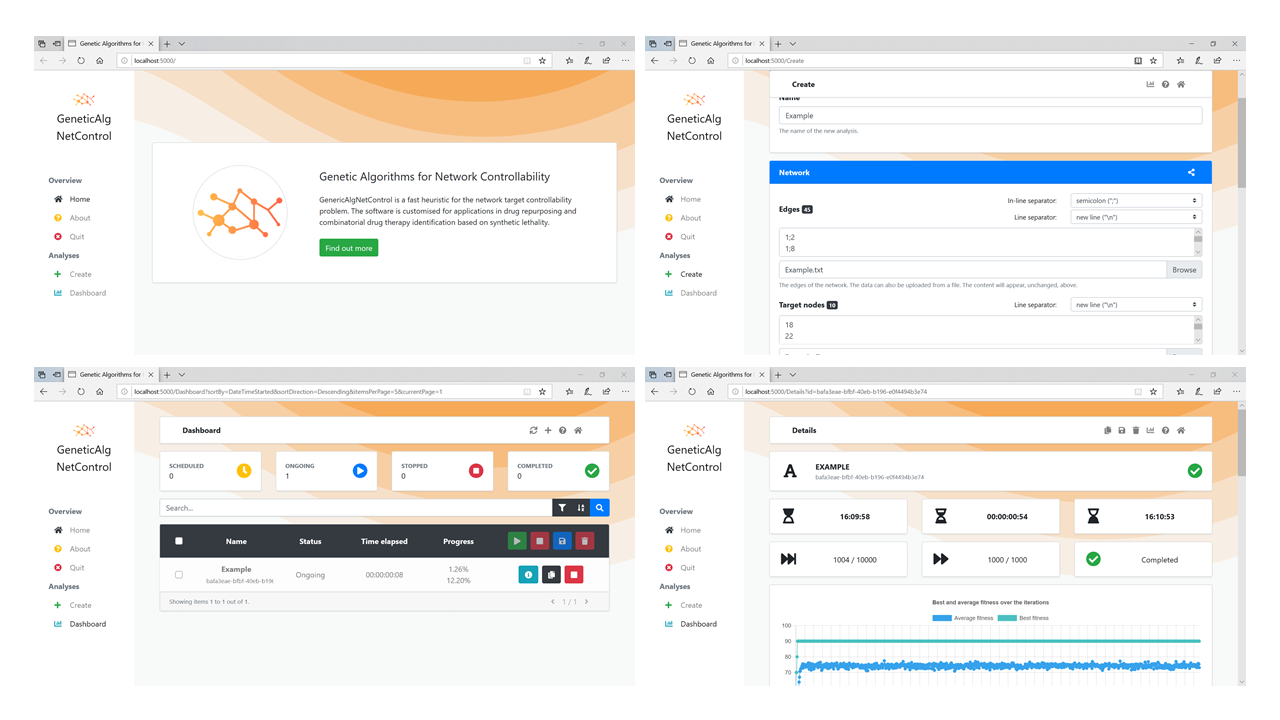}
    \end{center}
    \caption{The graphical user interface of the program. Top left: The start page. Top right: The form to create a new analysis. Bottom left: The dashboard containing the list of analyses. Bottom right: The details of an analysis.}
    \label{figure-screen}
\end{figure}

\section{Results}

\subsection{Benchmark data}

We applied the algorithm on several real world and randomly generated complex networks. The size of the networks varied from $32$ to over $3000$ nodes. An overview of the data sets is presented in Table \ref{table-data-sets}. We used the breast, pancreatic, and ovarian cancer cell line-specific protein-protein interaction networks documented in \cite{art-kcgp17}. We also used the breast, pancreatic, and ovarian cancer networks of \cite{art-kbskkm11}. We also considered several social interaction networks and electronic circuit networks documented in \cite{art-msikca02} and \cite{art-miklsasa04}. Finally, we generated several random graphs with the number of nodes ranging from 100 to 3000 and the edges distributed according to the Erd\"{o}s-R\'{e}nyi-, scale-free-, and small world-graph edge distributions, all of them generated using the Python networkx library (\cite{pro-hss08}). All networks are available as supplementary information.

\begin{table}[htb]
    \begin{center}
        \begin{tabular}{|p{.20\textwidth}|p{.20\textwidth}|p{.20\textwidth}|p{.10\textwidth}|p{.10\textwidth}|}
            \hline
            \textbf{Type} & \textbf{Network} & \textbf{Reference} & \textbf{Nodes} & \textbf{Edges} \\
            \hline
            \raggedright{Protein-protein interaction} & \raggedright{Breast DEF} & \raggedright{\cite{art-kcgp17}} & $1415$ & $2435$ \\
            \hline
            \raggedright{Protein-protein interaction} & \raggedright{Breast HCC1428} & \raggedright{\cite{art-kbskkm11}} & $1495$ & $2650$ \\
            \hline
            \raggedright{Protein-protein interaction} & \raggedright{Breast MDA-MB-361} & \raggedright{\cite{art-kbskkm11}} & $1478$ & $2590$ \\
            \hline
            \raggedright{Protein-protein interaction} & \raggedright{Ovarian DEF} & \raggedright{\cite{art-kcgp17}} & $1047$ & $1579$ \\
            \hline
            \raggedright{Protein-protein interaction} & \raggedright{Ovarian O1946} & \raggedright{\cite{art-kbskkm11}} & $1155$ & $1823$ \\
            \hline
            \raggedright{Protein-protein interaction} & \raggedright{Ovarian OVCA8} & \raggedright{\cite{art-kbskkm11}} & $1157$ & $1781$ \\
            \hline
            \raggedright{Protein-protein interaction} & \raggedright{Pancreatic AsPC-1} & \raggedright{\cite{art-kbskkm11}} & $1022$ & $1534$ \\
            \hline
            \raggedright{Protein-protein interaction} & \raggedright{Pancreatic DEF} & \raggedright{\cite{art-kcgp17}} & $991$ & $1484$ \\
            \hline
            \raggedright{Protein-protein interaction} & \raggedright{Pancreatic KP-3} & \raggedright{\cite{art-kbskkm11}} & $1134$ & $1757$ \\
            \hline
            \raggedright{Protein-protein interaction} & \raggedright{SIGNOR BrOvPa DEF} & \raggedright{\cite{art-kcgp17}} & $2913$ & $6729$ \\
            \hline
            \raggedright{Social} & \raggedright{Social Interaction 1} & \raggedright{\cite{art-miklsasa04}} & $67$ & $182$ \\
            \hline
            \raggedright{Social} & \raggedright{Social Interaction 3} & \raggedright{\cite{art-miklsasa04}} & $32$ & $96$ \\
            \hline
            \raggedright{Electronic circuit} & \raggedright{Electronic circuit 208} & \raggedright{\cite{art-msikca02}} & $122$ & $189$ \\
            \hline
            \raggedright{Electronic circuit} & \raggedright{Electronic circuit 420} & \raggedright{\cite{art-msikca02}} & $208$ & $189$ \\
            \hline
            \raggedright{Electronic circuit} & \raggedright{Electronic circuit 838} & \raggedright{\cite{art-msikca02}} & $512$ & $819$ \\
            \hline
            \raggedright{Erdős–Rényi} & \raggedright{Erdos-Renyi 100} & \raggedright{*} & $62$ & $47$ \\
            \hline
            \raggedright{Erdős–Rényi} & \raggedright{Erdos-Renyi 500} & \raggedright{*} & $497$ & $1270$ \\
            \hline
            \raggedright{Erdős–Rényi} & \raggedright{Erdos-Renyi 1000} & \raggedright{*} & $1000$ & $4952$ \\
            \hline
        \end{tabular}
    \end{center}
    \caption*{To be continued.}
\end{table}

\begin{table}[htb]
    \begin{center}
        \begin{tabular}{|p{.20\textwidth}|p{.20\textwidth}|p{.20\textwidth}|p{.10\textwidth}|p{.10\textwidth}|}
            \hline
            \textbf{Type} & \textbf{Network} & \textbf{Reference} & \textbf{Nodes} & \textbf{Edges} \\
            \hline
            \raggedright{Erdős–Rényi} & \raggedright{Erdos-Renyi 1500} & \raggedright{*} & $1500$ & $11258$ \\
            \hline
            \raggedright{Erdős–Rényi} & \raggedright{Erdos-Renyi 2000} & \raggedright{*} & $2000$ & $19869$ \\
            \hline
            \raggedright{Erdős–Rényi} & \raggedright{Erdos-Renyi 2500} & \raggedright{*} & $2500$ & $30976$ \\
            \hline
            \raggedright{Erdős–Rényi} & \raggedright{Erdos-Renyi 3000} & \raggedright{*} & $3000$ & $44713$ \\
            \hline
            \raggedright{Scale-free} & \raggedright{Scale Free 100} & \raggedright{**} & $100$ & $166$ \\
            \hline
            \raggedright{Scale-free} & \raggedright{Scale Free 500} & \raggedright{**} & $500$ & $816$ \\
            \hline
            \raggedright{Scale-free} & \raggedright{Scale Free 1000} & \raggedright{**} & $1000$ & $1793$ \\
            \hline
            \raggedright{Scale-free} & \raggedright{Scale Free 1500} & \raggedright{**} & $1500$ & $2526$ \\
            \hline
            \raggedright{Scale-free} & \raggedright{Scale Free 2000} & \raggedright{**} & $2000$ & $3401$ \\
            \hline
            \raggedright{Scale-free} & \raggedright{Scale Free 2500} & \raggedright{**} & $2500$ & $4316$ \\
            \hline
            \raggedright{Scale-free} & \raggedright{Scale Free 3000} & \raggedright{**} & $3000$ & $5236$ \\
            \hline
            \raggedright{Small world} & \raggedright{Small World 100} & \raggedright{***} & $100$ & $400$ \\
            \hline
            \raggedright{Small world} & \raggedright{Small World 500} & \raggedright{***} & $500$ & $2000$ \\
            \hline
            \raggedright{Small world} & \raggedright{Small World 1000} & \raggedright{***} & $1000$ & $4000$ \\
            \hline
            \raggedright{Small world} & \raggedright{Small World 1500} & \raggedright{***} & $1500$ & $6000$ \\
            \hline
            \raggedright{Small world} & \raggedright{Small World 2000} & \raggedright{***} & $2000$ & $8000$ \\
            \hline
            \raggedright{Small world} & \raggedright{Small World 2500} & \raggedright{***} & $2500$ & $10000$ \\
            \hline
            \raggedright{Small world} & \raggedright{Small World 3000} & \raggedright{***} & $3000$ & $12000$ \\
            \hline
        \end{tabular}
    \end{center}
    \caption{The data sets used for testing the algorithm (\cite{sof-p19}). * Generated in Python, using networkx.generators.random\_graphs.fast\_gnp\_random\_graph with $p = 0.005$. ** Generated in Python, using networkx.generators.directed.scale\_free\_graph with the default parameters. *** Generated in Python, using networkx.generators.random\_graphs.watts\_strogatz\_graph with $k = 4$ and $p = 0.2$. All isolated nodes were removed from the networks.}
    \label{table-data-sets}
\end{table}

For the protein-protein interaction networks we used as target nodes the cancer essential genes specific to each cell line, based on \cite{art-kbskkm11}. For the other networks, as target nodes we chose the top $5\%$ nodes with highest degree.

\subsection{The comparison setup}

All runs were executed with the same default values for the parameters of the algorithm, as detailed in Table \ref{table-parameters}. The algorithm was stopped after 1,000 generations with no improvement in the fitness of the best solution, up to a maximum of 10,000 generations. The results are presented in Table \ref{table-results} and in the supplementary data.

\begin{table}[htb]
    \begin{center}
        \begin{tabular}{|p{.25\textwidth}|p{.08\textwidth}|p{.08\textwidth}|p{.08\textwidth}|p{.08\textwidth}|p{.08\textwidth}|p{.08\textwidth}|}
            \hline
            \textbf{Network} & \textbf{N} & \textbf{E} & \textbf{T} & \textbf{I\textsubscript{ge}} & \textbf{I\textsubscript{gc}} & \textbf{I\textsubscript{gr}} \\
            \hline
            \raggedright{Breast DEF} & $1415$ & $2435$ & $112$ & $57$ & $74$ & $73$ \\
            \hline
            \raggedright{Breast HCC1428} & $1495$ & $2650$ & $126$ & $64$ & $82$ & $80$ \\
            \hline
            \raggedright{Breast MDA-MB-361} & $1478$ & $2590$ & $124$ & $61$ & $80$ & $77$ \\
            \hline
            \raggedright{Ovarian DEF} & $1047$ & $1579$ & $140$ & $92$ & $112$ & $111$ \\
            \hline
            \raggedright{Ovarian O1946} & $1155$ & $1823$ & $159$ & $108$ & $120$ & $120$ \\
            \hline
            \raggedright{Ovarian OVCA8} & $1157$ & $1781$ & $161$ & $99$ & $116$ & $115$ \\
            \hline
            \raggedright{Pancreatic AsPC-1} & $1022$ & $1534$ & $125$ & $78$ & $93$ & $91$ \\
            \hline
            \raggedright{Pancreatic DEF} & $991$ & $1484$ & $168$ & $109$ & $131$ & $131$ \\
            \hline
            \raggedright{Pancreatic KP-3} & $1134$ & $1757$ & $167$ & $103$ & $128$ & $126$ \\
            \hline
            \raggedright{SIGNOR BrOvPa DEF} & $2913$ & $6729$ & $145$ & $68$ & $90$ & $82$ \\
            \hline
            \raggedright{Social Interaction 1} & $67$ & $182$ & $14$ & $1$ & $1$ & $1$ \\
            \hline
            \raggedright{Social Interaction 3} & $32$ & $96$ & $7$ & $2$ & $3$ & $3$ \\
            \hline
            \raggedright{Electronic Circuit 208} & $122$ & $189$ & $25$ & $1$ & $3$ & $3$ \\
            \hline
            \raggedright{Electronic Circuit 420} & $208$ & $189$ & $42$ & $25$ & $25$ & $25$ \\
            \hline
            \raggedright{Electronic Circuit 838} & $512$ & $819$ & $103$ & $3$ & $15$ & $14$ \\
            \hline
            \raggedright{Erdos-Renyi 100} & $62$ * & $47$ & $5$ & $4$ & $4$ & $4$ \\
            \hline
            \raggedright{Erdos-Renyi 500} & $497$ * & $1270$ & $25$ & $1$ & $4$ & $4$ \\
            \hline
            \raggedright{Erdos-Renyi 1000} & $1000$ & $4952$ & $50$ & $1$ & $2$ & $1$ \\
            \hline
        \end{tabular}
    \end{center}
    \caption*{To be continued.}
\end{table}

\begin{table}[htb]
    \begin{center}
        \begin{tabular}{|p{.25\textwidth}|p{.08\textwidth}|p{.08\textwidth}|p{.08\textwidth}|p{.08\textwidth}|p{.08\textwidth}|p{.08\textwidth}|}
            \hline
            \textbf{Network} & \textbf{N} & \textbf{E} & \textbf{T} & \textbf{I\textsubscript{ge}} & \textbf{I\textsubscript{gc}} & \textbf{I\textsubscript{gr}} \\
            \hline
            \raggedright{Erdos-Renyi 1500} & $1500$ & $11258$ & $75$ & $2$ & $3$ & $1$ \\
            \hline
            \raggedright{Erdos-Renyi 2000} & $2000$ & $19869$ & $100$ & $2$ & $4$ & $1$ \\
            \hline
            \raggedright{Erdos-Renyi 2500} & $2500$ & $30976$ & $125$ & $3$ & $4$ & $1$ \\
            \hline
            \raggedright{Erdos-Renyi 3000} & $3000$ & $44713$ & $150$ & $3$ & $7$ & $1$ \\
            \hline
            \raggedright{Scale Free 100} & $100$ & $166$ & $5$ & $2$ & $2$ & $2$ \\
            \hline
            \raggedright{Scale Free 500} & $500$ & $816$ & $25$ & $17$ & $17$ & $17$ \\
            \hline
            \raggedright{Scale Free 1000} & $1000$ & $1793$ & $50$ & $37$ & $38$ & $38$ \\
            \hline
            \raggedright{Scale Free 1500} & $1500$ & $2526$ & $75$ & $60$ & $61$ & $61$ \\
            \hline
            \raggedright{Scale Free 2000} & $2000$ & $3401$ & $100$ & $71$ & $72$ & $72$ \\
            \hline
            \raggedright{Scale Free 2500} & $2500$ & $4316$ & $125$ & $93$ & $94$ & $94$ \\
            \hline
            \raggedright{Scale Free 3000} & $3000$ & $5236$ & $150$ & $94$ & $96$ & $96$ \\
            \hline
            \raggedright{Small World 100} & $100$ & $400$ & $5$ & $1$ & $1$ & $1$ \\
            \hline
            \raggedright{Small World 500} & $500$ & $2000$ & $25$ & $1$ & $1$ & $1$ \\
            \hline
            \raggedright{Small World 1000} & $1000$ & $4000$ & $50$ & $1$ & $3$ & $1$ \\
            \hline
            \raggedright{Small World 1500} & $1500$ & $6000$ & $75$ & $2$ & $5$ & $1$ \\
            \hline
            \raggedright{Small World 2000} & $2000$ & $8000$ & $100$ & $2$ & $7$ & $1$ \\
            \hline
            \raggedright{Small World 2500} & $2500$ & $10000$ & $125$ & $3$ & $9$ & $1$ \\
            \hline
            \raggedright{Small World 3000} & $3000$ & $12000$ & $150$ & $3$ & $12$ & $1$ \\
            \hline
        \end{tabular}
    \end{center}
    \caption{The results of the algorithm. \textbf{N}: the number of nodes in the network; \textbf{E}: the number of edges in the network; \textbf{T}: the number of target nodes in the network: \textbf{I}: the smallest number of input nodes found for controlling the control target set by the respective algorithm, i.e., \textbf{I\textsubscript{ge}} for the genetic algorithm, \textbf{I\textsubscript{gc}} for the constrained greedy algorithm, and \textbf{I\textsubscript{gr}} for the general (unconstrained) greedy algorithm. * The nodes without any inbound or outbound edge are not counted.}
    \label{table-results}
\end{table}

In addition, we applied the algorithm one more time on the protein-protein interaction networks, considering as preferred nodes the FDA approved drug-targets (\cite{art-wfglmgsjlsanilmgwcclpkw18}) existent in the network. The results are presented in Table \ref{table-results-drug} and in the supplementary data.

\begin{table}[htb]
    \begin{center}
        \begin{tabular}{|p{.15\textwidth}|p{.05\textwidth}|p{.05\textwidth}|p{.05\textwidth}|p{.05\textwidth}|p{.04\textwidth}|p{.05\textwidth}|p{.04\textwidth}|p{.05\textwidth}|p{.04\textwidth}|p{.05\textwidth}|}
            \hline
            \textbf{Network} & \textbf{N} & \textbf{E} & \textbf{T} & \textbf{P} & \textbf{I\textsubscript{ge}} & \textbf{IP\textsubscript{ge}} & \textbf{I\textsubscript{gc}} & \textbf{IP\textsubscript{gc}} & \textbf{I\textsubscript{gr}} & \textbf{IP\textsubscript{gr}} \\
            \hline
            \raggedright{Breast DEF (drug)} & $1415$ & $2435$ & $112$ & $123$ & $61$ & $13$ & $74$ & $6$ & $73$ & $5$ \\
            \hline
            \raggedright{Breast HCC1428 (drug)} & $1495$ & $2650$ & $126$ & $135$ & $63$ & $10$ & $82$ & $7$ & $80$ & $5$ \\
            \hline
            \raggedright{Breast MDA-MB-361 (drug)} & $1478$ & $2590$ & $124$ & $136$ & $63$ & $13$ & $80$ & $5$ & $77$ & $4$ \\
            \hline
            \raggedright{Ovarian DEF (drug)} & $1047$ & $1579$ & $140$ & $100$ & $94$ & $10$ & $112$ & $7$ & $111$ & $6$ \\
            \hline
            \raggedright{Ovarian O1946 (drug)} & $1155$ & $1823$ & $159$ & $104$ & $108$ & $18$ & $120$ & $10$ & $119$ & $11$ \\
            \hline
            \raggedright{Ovarian OVCA8 (drug)} & $1157$ & $1781$ & $161$ & $105$ & $101$ & $15$ & $115$ & $11$ & $115$ & $10$ \\
            \hline
            \raggedright{Pancreatic AsPC-1 (drug)} & $1022$ & $1534$ & $125$ & $90$ & $73$ & $7$ & $93$ & $5$ & $91$ & $4$ \\
            \hline
            \raggedright{Pancreatic DEF (drug)} & $991$ & $1484$ & $168$ & $86$ & $120$ & $18$ & $131$ & $10$ & $131$ & $9$ \\
            \hline
            \raggedright{Pancreatic KP-3 (drug)} & $1134$ & $1757$ & $167$ & $94$ & $104$ & $15$ & $128$ & $8$ & $126$ & $8$ \\
            \hline
            \raggedright{SIGNOR BrOvPa DEF (drug)} & $2913$ & $6729$ & $201$ & $145$ & $80$ & $19$ & $91$ & $4$ & $82$ & $1$ \\
            \hline
        \end{tabular}
    \end{center}
    \caption{The results of the algorithm.         \textbf{N}: the number of nodes in the network; \textbf{E}: the number of edges in the network; \textbf{T}: the number of control target nodes in the network; \textbf{P}: the number of preferred (i.e., drug-targetable) nodes in the network; \textbf{I}: the smallest number of input nodes found for controlling the control target set; \textbf{IP} the number of preferred input nodes in the best solution; \textbf{{ge}}: the results for the genetic algorithm; \textbf{{gc}}: the results for the constrained greedy algorithm;\textbf{{gr}}: the results for the general (unconstrained) greedy algorithm.}
    \label{table-results-drug}
\end{table}

We compared the results of our genetic algorithm to the results of the greedy algorithm described in \cite{art-cgkkp18}, applied on the same data sets. To make the comparison possible, we limited both algorithms to running for a maximum of 10,000 total iterations (translated to 10,000 generations for the genetic algorithm), stopping if there was no improvement in the best result over the past 1,000 iterations / generations. To investigate the effect of the limited length pathways in our algorithm, we ran the greedy algorithm in two different settings: with the control path's length upper bounded by the same parameter as in the genetic algorithm, and with it unconstrained.

The three algorithms work quite differently, not only in their inner logic for optimising the objective, but also in their output. The genetic algorithm maintains in each step of its search a family of solutions, some better than others (from the point of view of the optimisation problem to be solved), but all valid solutions in terms of controlling the given set of targets. Thus, each run of the genetic algorithm offers as an output several different solutions. This is in contrast with the greedy algorithms, where only one solution is found in one run and multiple runs have to be done to collect multiple solutions (of variable optimisation quality).

\subsection{The comparison results}

The first benchmark objective we compared against was the size of the smallest set of input nodes found by each of the three algorithms, with the smallest being the best. The results are presented in Table \ref{table-results}. 

\begin{figure}[htb]
    \begin{center}
        \includegraphics[width=0.9\textwidth]{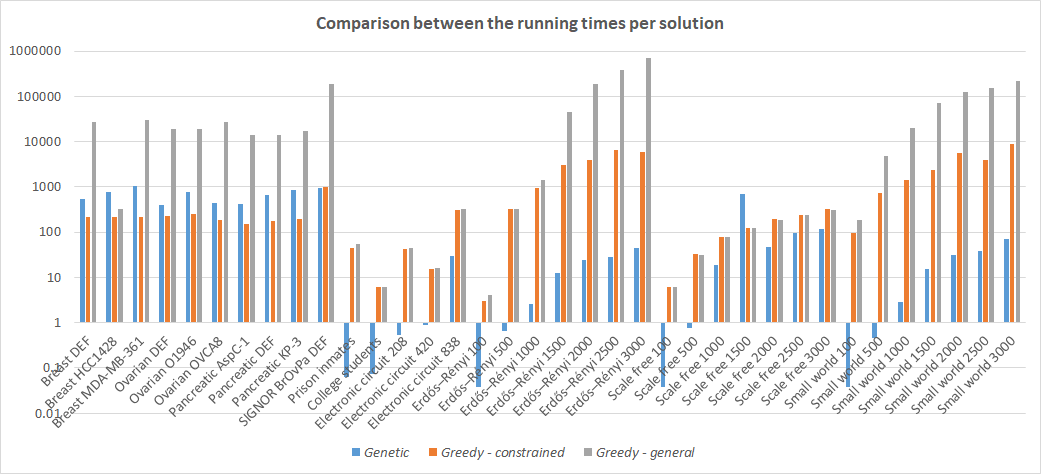}
    \end{center}
    \caption{A comparison between the results of the three algorithms: the running time per solution. The data is displayed on logarithmic scale.}
    \label{figure-comparison-standard-running-time-per-solution}
\end{figure}

\begin{figure}[htb]
    \begin{center}
        \includegraphics[width=0.9\textwidth]{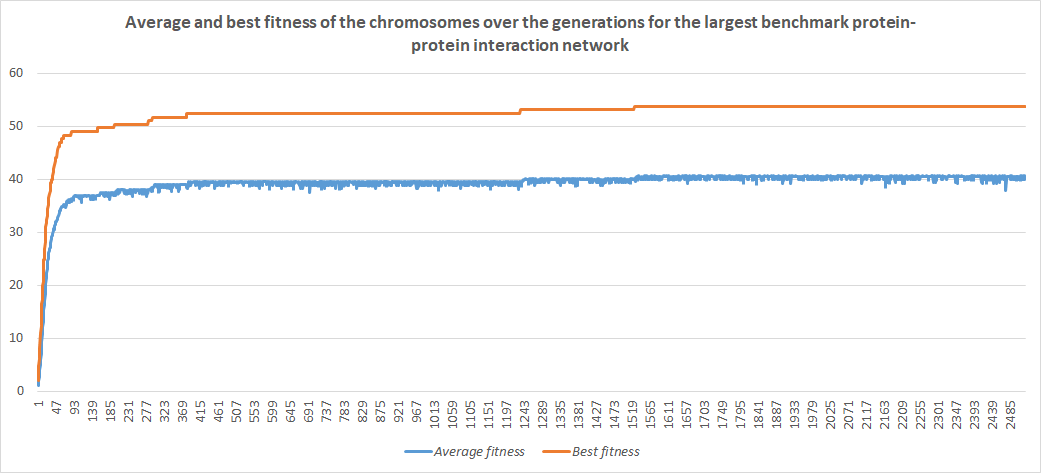}
    \end{center}
    \caption{The average and best fitness of the chromosomes over the generations in the case of the largest benchmark protein-protein interaction network.}
    \label{figure-convergence-speed}
\end{figure}

We also compared the running time required by the algorithms to complete on each of the benchmark networks and the speed of convergence towards a good solution. The results, reported as running time per solution and convergence speed are in Figures \ref{figure-comparison-standard-running-time-per-solution} and \ref{figure-convergence-speed}.

Finally, we compared the ability of the three algorithms to maximise the use of preferred nodes. We did this on the biological networks, with the preferred nodes being the set of FDA-approved drug targets. We also did a literature-based validation of the relevance of the results found by the genetic algorithm in each of the cancer networks. The results are in Tables \ref{table-results-drug}, \ref{table-validation-proteins} and \ref{table-validation-drugs}.

\begin{table}[htb]
    \begin{center}
        \begin{tabular}{|p{.25\textwidth}|p{.25\textwidth}|p{.25\textwidth}|}
            \hline
            \textbf{Breast} & \textbf{Ovarian} & \textbf{Pancreatic} \\
            \hline
            \textit{RET} (5) \newline \textbf{EGFR} (6) \newline \textit{MTCP1} (5) \newline \textbf{AURKB} (6) \newline \textbf{RAF1} (5) \newline \textbf{SRC} (5) \newline PLRG1 (2) & \textbf{TGFB1} (5) \newline \textbf{SRC} (5) \newline \textit{SET} (5) \newline \textit{LCK} (5) \newline \textbf{MAPK3} (6) \newline \textit{CSNK2A1} (5) & \textbf{IGF1R} (6) \newline \textbf{ERBB4} (5) \newline \textbf{KRAS} (6) \newline \textit{ABL1} (6) \newline \textit{PDPK1} (5) \newline PPP2R1A (5) \newline DUSP7 (5) \newline \textbf{ERBB2} (5) \newline \textit{CSNK2A1} (5) \newline \textit{CDK1} (6) \newline \textbf{MTOR} (6) \\
            \hline
        \end{tabular}
    \end{center}
    \caption{The proteins that control the most target proteins, for each cancer type. In bold letters: proteins that are known to be of significance in the corresponding cancer type. In italic letters: proteins that are known to be of significance in other cancer types. In brackets: number of target proteins controlled by the protein by itself (the sets of target proteins controlled by different proteins might overlap).}
    \label{table-validation-proteins}
\end{table}

\begin{table}[htb]
    \begin{center}
        \begin{tabular}{|p{.25\textwidth}|p{.25\textwidth}|p{.25\textwidth}|}
            \hline
            \textbf{Breast} & \textbf{Ovarian} & \textbf{Pancreatic} \\
            \hline
            \textit{amuvatinib} \newline \textit{cabozantinib} \newline \textbf{canertinib} \newline \textbf{dasatinib} \newline fostamatinib \newline \textbf{KX-01} \newline \textbf{lapatinib} \newline \textit{LErafAON} \newline \textit{ponatinib} \newline \textbf{selpercatinib} \newline \textit{sorafenib} \newline \textit{sunitinib} \newline \textbf{vandetanib} \newline \textbf{varlitinib} \newline \textit{XL281} & \textbf{dasatinib} \newline fostamatinib \newline \textit{nintedanib} \newline \textit{ponatinib} \newline \textit{seliciclib} \newline \textbf{ulixertinib} \newline \textit{XL228} & \textit{AT-7519} \newline \textit{brigatinib} \newline \textbf{celecoxib} \newline fostamatinib \newline \textbf{linsitinib} \newline \textit{rhIGFBP-3} \newline \textit{ridaforolimux} \textit{seliciclib} \newline \textit{SF1126} \newline \textit{XL228} \newline \textit{XL765} \\
            \hline
        \end{tabular}
    \end{center}
    \caption{The drugs that target the identified control proteins. In bold letters: drugs approved or under investigation for usage in the treatment of the corresponding cancer type. In italic letters: drugs under investigation for treatment of other or unspecified cancer or tumour types.}
    \label{table-validation-drugs}
\end{table}

\section{Discussion}

We applied the algorithm on a series of networks ranging from a few tens of nodes, to several thousands of them. For each such network size, we analysed several network types (real-life, random, scale-free, small world), each with a varying number of edges and target nodes. The running times ranged from a few seconds on the smaller networks, up to several hours on the bigger ones.

\subsection*{Size of the best solutions}

\begin{figure}[htb]
    \begin{center}
        \includegraphics[width=0.9\textwidth]{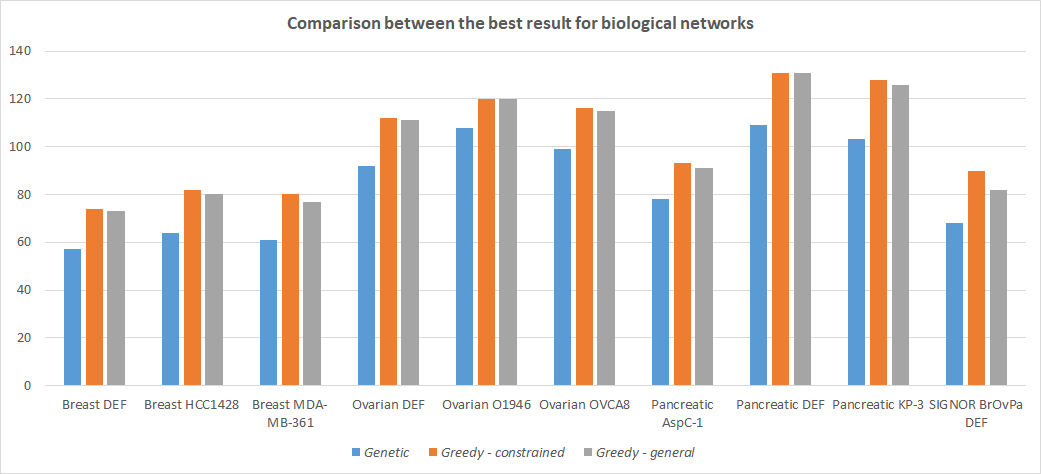}
    \end{center}
    \caption{A comparison between the results of the three algorithms: the number of input nodes controlling the set of disease-specific essential genes in each of the biological benchmark networks.}
    \label{figure-comparison-standard-result}
\end{figure}

In the case of the cancer protein-protein interaction networks, the genetic algorithm returned input sets of size $10 - 25\%$ smaller than the constrained greedy algorithm and $10 - 20\%$ smaller than the general greedy algorithm. The comparison can be seen in Figure \ref{figure-comparison-standard-result}.

In the case of the non-biological networks, the size of the smallest input sets is virtually identical in the three algorithms, even though the sets themselves may be different.

An interesting aspect can be seen in the analyses of the well-connected Erd\"{o}s-R\'{e}nyi and small world random networks, where the genetic algorithm succeeds in finding the optimal solution. To see this, let's consider the case of the largest small world benchmark network. The maximum path length being set at $50$ means that a given input node can control at most $51$ nodes in the network (itself, and $50$ others, through consecutive edges). Thus, for a set of $150$ target nodes, there have to be at least $\left[ \frac{150}{51} \right] + 1 = 3$ input nodes. The genetic algorithm indeed identifies exactly $3$ input nodes, with a maximum length of the control path of $49$. The general greedy algorithm, being allowed to use longer control paths, identified one input node that controls the entire target set, at the cost of an increased maximum path length of $149$ and a running time $40$ times longer.

\subsection*{Running times and convergence speed}

A big difference in the results was in the running time of the algorithms to complete. The genetic algorithm was the fastest algorithm in the random Erd\"{o}s-R\'{e}nyi networks, in the random small world networks, and in some of the real-life non-biological networks. For the biological and the scale-free networks the fastest was the constrained greedy algorithm, with the genetic algorithm the slowest of the three, and in general 2-4 times slower than the general greedy algorithm. However, the genetic algorithm provides simultaneously several solutions (in our benchmarks tests we set the population size parameter to 80), a key advantage of this method. Because of this, the comparison of the running times per solution shows the genetic algorithm to be the fastest of the three in all except a handful of examples, see Figure \ref{figure-comparison-standard-running-time-per-solution}. Compared to the general greedy algorithm, the running times per solution of the genetic algorithm were up to 10,000 times faster, except in the case of two networks (one biological, the other a random scale-free) where the greedy algorithm seemed to stumble almost immediately on a solution. Compared to the constrained greedy algorithm, the genetic algorithm was 10-5,000 times faster per solution, except in the case of the the biological networks (where it was about 2-4 times slower) and the same scale-free random network where the generic greedy also performed unusually well.

The setup in which we ran the genetic algorithm was to search thoroughly for a good solution through 10,000 generations. We compared the evolution of the quality of the solutions throughout the generations and we noticed that a good solution (i.e., a solution with the fitness within $10\%$ of the solution obtained after 10,000 generations) is in fact achieved very quickly, typically within a few tens of generations from the start. Figure \ref{figure-convergence-speed} illustrates the average and best fitness of the chromosomes in each generation of the algorithm on the largest protein-protein interaction network in the data set. This suggests that the genetic algorithm may be applied successfully with a much lower number of generations, perhaps as low as 100, adding a considerable speed-up to it.

\subsection*{Maximising the number of preferred nodes}

\begin{figure}[htb]
    \begin{center}
        \includegraphics[width=0.9\textwidth]{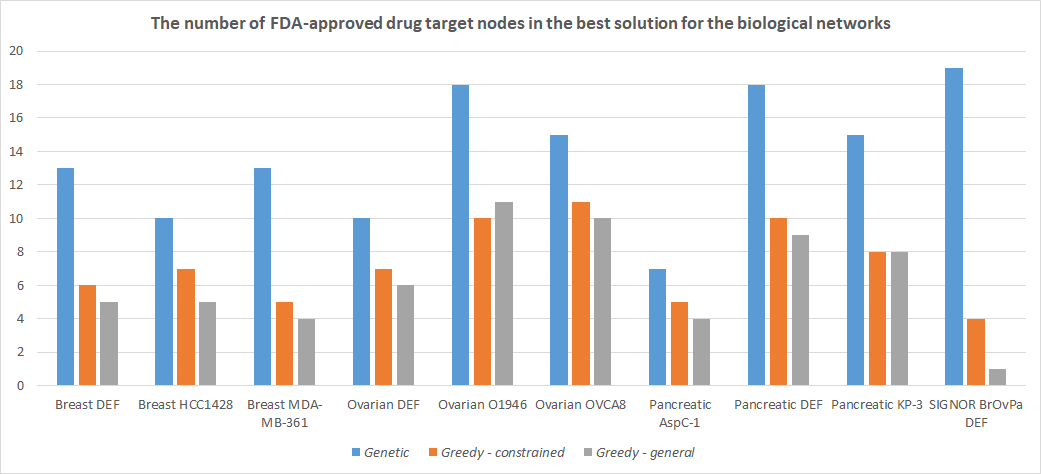}
    \end{center}
    \caption{The number of preferred (FDA-approved drug targets) nodes in the best solution found by each of the three algorithms.}
    \label{figure-comparison-preferred-drug-target-number}
\end{figure}

\begin{figure}[htb]
    \begin{center}
        \includegraphics[width=0.9\textwidth]{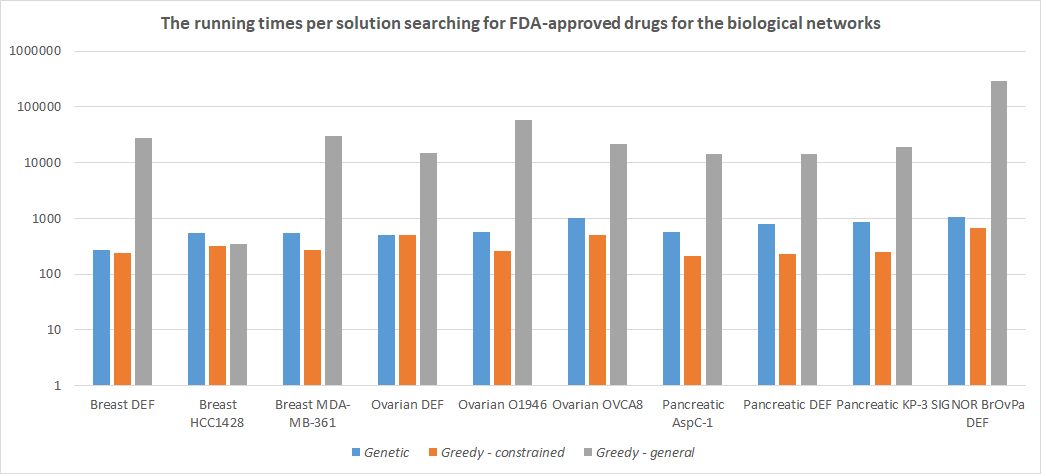}
    \end{center}
    \caption{The runtime for finding preferred (FDA-approved drug targets) nodes in the best solution found by each of the three algorithms. The data is displayed on logarithmic scale.}
    \label{figure-comparison-preferred-running-time}
\end{figure}

\begin{figure}[htb]
    \begin{center}
        \includegraphics[width=0.9\textwidth]{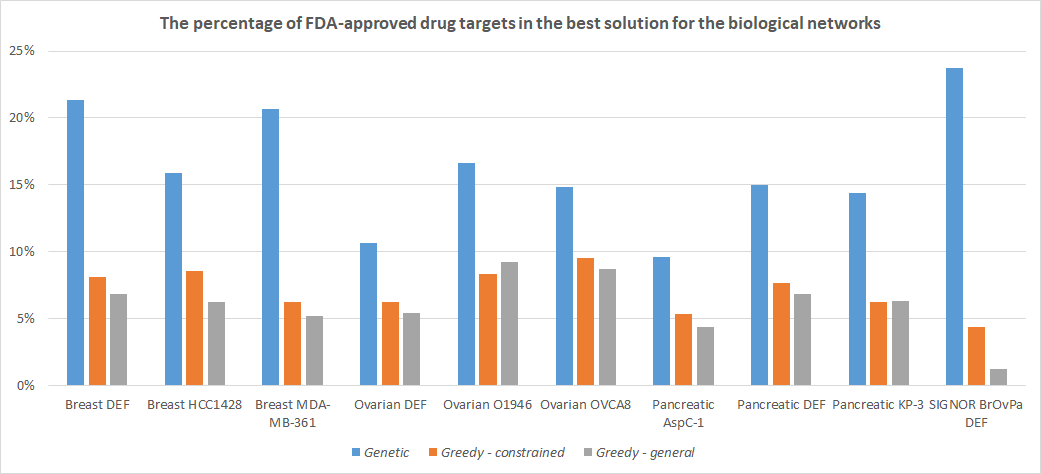}
    \end{center}
    \caption{The percentage of preferred (FDA-approved drug targets) nodes in the best solution found by each of the three algorithms.}
    \label{figure-comparison-preferred-drug-target-percentage}
\end{figure}

\begin{figure}[htb]
    \begin{center}
        \includegraphics[width=0.9\textwidth]{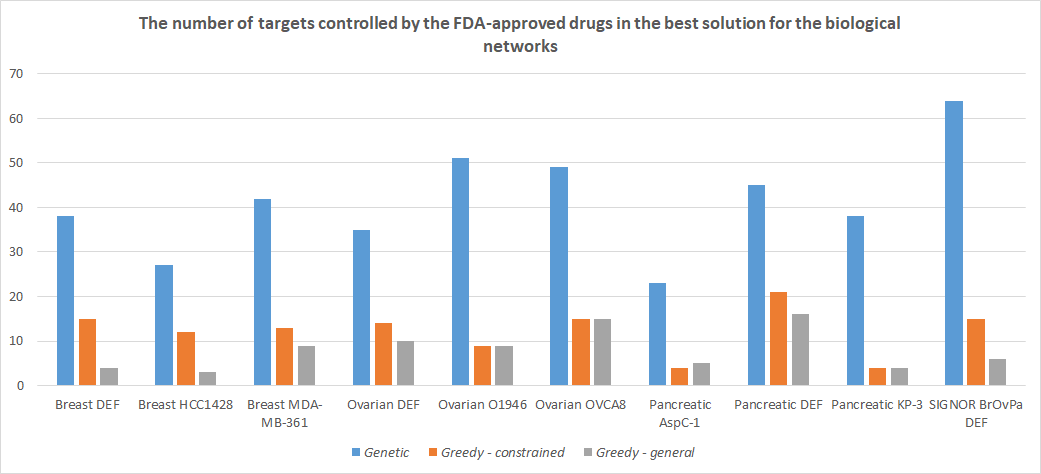}
    \end{center}
    \caption{The number of essential genes controlled by the preferred (FDA-approved drug targets) nodes in the best solution found by each of the three algorithms.}
    \label{figure-comparison-preferred-targets-controlled}
\end{figure}

We applied the three algorithms on the biological networks with the additional optimisation objective of maximising the selection as input nodes of FDA-approved drug targets in the network (preferred nodes). In all cases, the sets of input nodes returned by the genetic algorithm contained more preferred nodes than the ones returned by the other algorithms (Figure \ref{figure-comparison-preferred-drug-target-number}), with a running time per solution similar to that of the constrained greedy and better than that of the general greedy algorithm (Figure \ref{figure-comparison-preferred-running-time}). Even more, the percentage of the preferred nodes relative to the size of the input nodes in the best solution was in general two to four times higher in the case of the genetic algorithm (Figure \ref{figure-comparison-preferred-drug-target-percentage}). This led to more control target nodes being controlled by preferred nodes (Figure \ref{figure-comparison-preferred-targets-controlled}), i.e., leading to predictions of potentially more efficient drugs. This has as a consequence a clear improvement in the applicability of the algorithm in the biomedical domain for drug repurposing, an aspect that we discuss next.

\subsection*{Therapeutically-relevant findings}

We analysed in details the FDA-approved drug targets predicted by our algorithm to control the most essential genes in each of our network. The results are presented in Table \ref{table-validation-proteins}. We used the public DrugBank database \cite{art-wfglmgsjlsanilmgwcclpkw18} to find  drugs targeting the proteins in Table \ref{table-validation-proteins} and known to be used in cancer therapeutics. The results are presented in the Table \ref{table-validation-drugs}.

Out of the seven top controlling proteins for the analysed breast cancer networks, four proteins are known to be of significance in breast cancer proliferation: \textit{EGFR} (\cite{art-bgo10, art-m02, art-cd94}), \textit{AURKB} (\cite{art-gzcyvwyl10}), \textit{RAF1} (\cite{art-lgzlsyy09, art-mtldclrkjdg06, art-cnkpj95}) and \textit{SRC} (\cite{art-pm09, art-mnh08, art-hn08}). These proteins are targetted by several FDA-approved and investigational drugs used in fighting breast cancer (\textit{lapatinib}, \textit{vandetanib}, \textit{canertinib}, \textit{varlitinib}, \textit{KX-01}, \textit{dasatinib}). Additionally, two other drugs targetting the same proteins are investigated for use in treating unspecified cancer types (\textit{LErafAON}, \textit{XL281}). Two of the other top controlling proteins, \textit{RET} and \textit{MTCP1}, are known to be significant in other types of cancer, such as ovarian cancer (\cite{art-glxpztzczzjjwl20}), pancreatic cancer (\cite{art-dh09}), prostate cancer (\cite{art-vrmbsslctlplpyhcwmcned20}) or brain cancer (\cite{art-hlwl18}). Even more, there already exists an FDA-approved breast cancer drug targeting these proteins (\textit{selpercatinib}), as well as several approved and investigational drugs used for treating other cancer types (\textit{ponatinib}, \textit{sorafenib}, \textit{sunitinib}, \textit{cabozantinib}, \textit{amuvatinib}). In particular, the drug \textit{sorafenib}, used for the treatment of kidney and liver cancers, targets two of our top four controlling proteins, which may indicate its potential use in breast cancer as well. Indeed, there are several completed clinical trials researching the drug in the treatment of breast cancer (\cite{tri-chmc14, tri-nci06}). Additionally, the drug \textit{cabozantinib} is also on trial for breast cancer treatment (\cite{tri-mgh19, tri-dfci16}). Among the top controlling proteins was also \textit{PLRG1}, for which no drug exists, marking it as a potential drug-target for future research.

For the analysed ovarian cancer networks, the algorithm identified six top controlling proteins. Three of these are documented as being of significance in ovarian cancer: \textit{TGFB1} (\cite{art-sberch09, art-ejbl00}), \textit{SRC} (\cite{art-lwjtf10, art-hlthlkkmcgswsmdsgs06, art-wweptdbmg00}) and \textit{MAPK3} (\cite{art-slmqlaylc20, art-ywt20, art-cympp00}), with several drugs targeting these proteins already approved or being under investigation for various types of cancer treatment (\textit{dasatinib}, \textit{XL228}, \textit{seliciclib}, \textit{ulixertinib}, \textit{ponatinib}, \textit{nintedanib}). Furthermore, the other three top controlling proteins, \textit{SET}, \textit{LCK} and \textit{CSNK2A1}, are significant in other cancer types, such as leukemia (\cite{art-hb06, art-jfwkgdnf08}), colorectal cancer (\cite{art-ctrspczlsmrg19, art-fsws18}) and prostate cancer (\cite{art-awa06}).

For the pancreatic cancer networks analysed by the algorithm, we identified eleven top controlling proteins. Five of them are known to be significant in pancreatic cancer: \textit{IGF1R} (\cite{art-dsmdkdrpmhbybdtakncfchfokhwn20, art-zwyzjxxyfwchzlywnvqd20, art-zdnlpm03}), \textit{ERBB4} (\cite{art-halbssd20}), \textit{KRAS} (\cite{art-cc20, art-stscdzwawexmgztmcvpeskcc20, art-dmusggg06}), \textit{ERBB2} (\cite{art-bpd20, art-samsubwbbabp05, art-npvkmj01}) and \textit{MTOR} (\cite{art-lfcyqzclxyzz20, art-zzjwz20, art-hlm08}) and targeted by several FDA-approved and investigational cancer drugs (\textit{XL228}, \textit{rhIGFB-3}, \textit{linsitinib}, \textit{brigatinib}, \textit{SF1126}, \textit{XL765}, \textit{ridaforolimus}). Four of the other identified top controlling proteins, \textit{ABL1}, \textit{PDPK1}, \textit{CSNK2A1} and \textit{CDK1} are documented as significant in other types of cancer, for example breast cancer (\cite{art-jbwnrsc10, art-ssp08}) and prostate cancer (\cite{art-awa06, art-xnsrnakb09}). They are, as well, targeted by several drugs under investigation for treating cancer (\textit{AT-7519}, \textit{seliciclib}). In particular, the drug \textit{brigatinib} could be especially powerful in this case, as it targets four of the eleven top controlling proteins (\textit{ABL1}, \textit{IGF1R}, \textit{ERBB4}, \textit{ERBB2}). Another interesting case is the non-steroidal anti-inflammatory drug \textit{celecoxib} targeting \textit{PDPK1}, used to manage symptoms of arthritis pain and in familial adenomatous polyposis, which has also been under investigation as potential cancer chemo-preventive and therapeutic drug \cite{art-gtbgak12}. Indeed, there are several ongoing and completed trials researching the effect of the drug in pancreatic cancer (\cite{tri-sahsmzu23, tri-mdacc05}). Among the top controlling proteins were also \textit{PPP2R1A} and \textit{DUSP7}, for which no drug exists, marking them as well as potential drug-targets for future research.

Furthermore, we found the drug \textit{fostamatinib}, used for the treatment of rheumatoid arthritis and immune thrombocytopenic purpura, which targets five out of the seven top controlling proteins for breast cancer, four out of the six for ovarian cancer, and seven out of the eleven for pancreatic cancer. Our algorithm thus suggests that the drug could potentially be used in cancer treatment. This idea is supported by several completed clinical trials for using \textit{fostamatinib} in treating lymphoma (\cite{tri-a13, tri-rp10}) and one ongoing trial for ovarian cancer (\cite{tri-skcccjh20}).

\subsection*{Conclusions}

In this paper we proposed a new  solution for the target network controllability problem. Our search strategy is based on a genetic algorithm, where the population in each generation of the training of the algorithm is a set of valid solutions to the network controllability problem. The algorithm turns out to be scalable, with its performance staying strong even for very large networks. 

The number of edges in a network alone does not seem to influence the performance of the algorithm. The increase in the running time as the network size increases, is mainly caused by the increasing number of nodes and target nodes. This makes the genetic algorithm optimally suited to be applied on very large networks, where many solutions need to be collected. 

The genetic algorithm provides, at every step, a family of solutions, while the greedy algorithms offer only one. This is a key advantage of this algorithm, especially for drug repurposing applications, where multiple alternative solutions are important to collect and compare. 

The genetic algorithm comes by design with a set-limit on the maximum length of control paths from the input to the target nodes. This is a feature that is of particularly important interest in applications in medicine, where the effects of a drug dissipate quickly over longer signalling paths. The focused search upstream of the target nodes led to the genetic algorithm drastically improving the percentage of FDA-approved drug targets selected in its solution, a clear step forward towards applications in combinatorial drug selection and drug repurposing. 
The drugs identified by our algorithm as potentially efficient for breast, ovarian, and pancreatic cancer correlate well with recent literature results, and some of our suggestions have already been subject to several clinical studies. This strengthens the potential of our approach for studies in synthetic lethality-driven drug repurposing.

\section*{Author Contributions}

V.P., I.N., E.C. and I.P. conceived and designed the study. V.P. planned and coded the implementation, and performed the numerical calculations. K.K. collected the data for the protein-protein interaction networks and generated the networks. V.P. and I.P. analysed the results. All authors contributed to writing the article and approved its final form. 

\section*{Conflict of Interest Statement}

The authors declare that the research was conducted in the absence of any commercial or financial relationships that could be construed as a potential conflict of interest.

\section*{Funding}

This work was partially supported by the Academy of Finland (project 311371/2017) and by the Romanian National Authority for Scientific Research and Innovation (POC grant P\_37\_257 and PED grant 2391).

\bibliographystyle{plain}
\bibliography{bibliography}

\begin{thebibliography}{10}

\bibitem{art-awa06}
Kashif~A. Ahmad, Guixia Wang, and Khalil Ahmed.
\newblock Intracellular hydrogen peroxide production is an upstream event in
  apoptosis induced by down-regulation of casein kinase 2 in prostate cancer
  cells.
\newblock {\em Molecular cancer research}, 4(5):331--338, 2006.

\bibitem{tri-a13}
{AstraZeneca}.
\newblock Study to learn if 200mg test drug (fostamatinib) helps people with
  large b-cell lymphoma,a type of blood cancer, 2013.
\newblock Available at \url{https://clinicaltrials.gov/ct2/show/NCT01499303}.

\bibitem{Barabasi:2011aa}
Albert-L{\'a}szl{\'o} Barab{\'a}si, Natali Gulbahce, and Joseph Loscalzo.
\newblock Network medicine: a network-based approach to human disease.
\newblock {\em Nature reviews. Genetics}, 12(1):56--68, 01 2011.

\bibitem{art-bpd20}
Laurence Booth, Andrew Poklepovic, and Paul Dent.
\newblock Neratinib decreases pro-survival responses of [sorafenib +
  vorinostat] in pancreatic cancer.
\newblock {\em Biochemical pharmacology}, 178:114067, 2020.

\bibitem{art-bgo10}
Monika~L. Burness, Tatyana~A. Grushko, and Olufunmilayo~I. Olopade.
\newblock Epidermal growth factor receptor in triple-negative and basal-like
  breast cancer: Promising clinical target or only a marker?
\newblock {\em Cancer journal}, 16(1):23--32, 2010.

\bibitem{art-cnkpj95}
L.S. Callans, H.~Naama, M.~Khandelwal, R.~Plotkin, and L.~Jardines.
\newblock Raf-1 protein expression in human breast cancer cells.
\newblock {\em Annals of surgical oncology}, 2:38--42, 1995.

\bibitem{Cheng:2019aa}
Feixiong Cheng, Istv\'{a}n~A. Kov\'{a}cs, and Albert-L\'{a}szl\'{o}
  Barab\'{a}si.
\newblock Network-based prediction of drug combinations.
\newblock {\em Nature Communications}, 10(1):1197, 2019.

\bibitem{art-cc20}
Jiongjia Cheng and John~R. Cashman.
\newblock Pawi-2 overcomes tumor stemness and drug resistance via cell cycle
  arrest in integrin b 3-kras-dependent pancreatic cancer stem cells.
\newblock {\em Scientific reports}, 10(1):9162, 2020.

\bibitem{art-cd94}
S.A. Chrysogelos and R.B. Dickson.
\newblock Egf receptor expression, regulation, and function in breast cancer.
\newblock {\em Breast cancer research and treatment}, 29(1):29--40, 1994.

\bibitem{tri-chmc14}
{City of Hope Medical Center}.
\newblock Sorafenib and vinorelbine in treating women with stage iv breast
  cancer, 2014.
\newblock Available at \url{https://clinicaltrials.gov/ct2/show/NCT00828074}.

\bibitem{art-ctrspczlsmrg19}
Ion Crist\'{o}bal, Blanca Torrej\'{o}n, Jaime Rubio, Andrea Santos, Manuel
  Pedregal, Cristina Caram{\'e}s, Sandra Zazo, Melani Luque, Marta
  Sanz-Alvarez, Juan Madoz-G\'{u}rpide, Federico Rojo, and Jes\'{u}s
  Garc\'{i}a-Foncillas.
\newblock Deregulation of set is associated with tumor progression and predicts
  adverse outcome in patients with early-stage colorectal cancer.
\newblock {\em Journal of clinical medicine}, 8(3), 2019.

\bibitem{art-cympp00}
W.~Cui, E.~M. Yazlovitskaya, M.~S. Mayo, J.~C. Pelling, and D.~L. Persons.
\newblock Cisplatin-induced response of c-jun n-terminal kinase 1 and
  extracellular signal--regulated protein kinases 1 and 2 in a series of
  cisplatin-resistant ovarian carcinoma cell lines.
\newblock {\em Molecular carcinogenesis}, 29(4):219--228, 2000.

\bibitem{art-cgkkp18}
Eugen Czeizler, Cristian Gratie, Wu~Kai~Chiu, Krishna Kanhaiya, and Ion Petre.
\newblock Structural target controllability of linear networks.
\newblock {\em IEEE/ACM Transactions on Computational Biology and
  Bioinformatics}, 15:1217--1228, 2018.

\bibitem{tri-dfci16}
{Dana-Farber Cancer Institute}.
\newblock Cabozantinib for metastatic triple negative brca, 2016.
\newblock Available at \url{https://clinicaltrials.gov/ct2/show/NCT01738438}.

\bibitem{Davis:2017aa}
Mark~M Davis, Cristina~M Tato, and David Furman.
\newblock Systems immunology: just getting started.
\newblock {\em Nature immunology}, 18(7):725--732, 06 2017.

\bibitem{art-dh09}
Timothy~R. Donahue and O.~Joe Hines.
\newblock Cxcr2 and ret single nucleotide polymorphisms in pancreatic cancer.
\newblock {\em World journal of surgery}, 33(4):710--715, 2009.

\bibitem{art-dmusggg06}
Ute Dreissigacker, Meike~S. Mueller, Monika Unger, Patrizia Siegert, Felicitas
  Genze, Peter Gierschik, and Klaudia Giehl.
\newblock Oncogenic k-ras down-regulates rac1 and rhoa activity and enhances
  migration and invasion of pancreatic carcinoma cells through activation of
  p38.
\newblock {\em Cell signalling}, 18(8):1156--1168, 2006.

\bibitem{art-dsmdkdrpmhbybdtakncfchfokhwn20}
Chunxia Du, Annacarolina da~Silva, Vicente Morales-Oyarvide, Andressa
  Dias~Costa, Margaret~M. Kozak, Richard~F. Dunne, Douglas~A. Rubinson,
  Kimberly Perez, Yohei Masugi, Tsuyoshi Hamada, Lauren~K. Brais, Chen Yuan,
  Ana Babic, Matthew~D. Ducar, Aaron~R. Thorner, Andrew Aguirre, Matthew~H.
  Kulke, Kimmie Ng, Thomas~E. Clancy, Jennifer~J. Findeis-Hosey, Daniel~T.
  Chang, Jason~L. Hornick, Charles~S. Fuchs, Shuji Ogino, Albert~C. Koong,
  Aram~F. Hezel, Brian~M. Wolpin, and Jonathan~A. Nowak.
\newblock Insulin-like growth factor-1 receptor expression and disease
  recurrence and survival in patients with resected pancreatic ductal
  adenocarcinoma.
\newblock {\em Cancer Epidemiology and Prevention Biomarkers}, 2020.

\bibitem{art-ejbl00}
A.~Evangelou, S.~K. Jindal, T.~J. Brown, and M.~Letarte.
\newblock Down-regulation of transforming growth factor beta receptors by
  androgen in ovarian cancer cells.
\newblock {\em Cancer research}, 60(14):929--935, 2000.

\bibitem{art-fsws18}
Hirota Fujiki, Eisaburo Sueoka, Tatsuro Watanabe, and Masami Suganuma.
\newblock The concept of the okadaic acid class of tumor promoters is revived
  in endogenous protein inhibitors of protein phosphatase 2a, set and cip2a, in
  human cancers.
\newblock {\em Journal of cancer research and clinical oncology},
  144(12):2339--2349, 2018.

\bibitem{Gao:2014aa}
Jianxi Gao, Yang-Yu Liu, Raissa~M. D'Souza, and Albert-L{\'a}szl{\'o}
  Barab{\'a}si.
\newblock Target control of complex networks.
\newblock {\em Nature Communications}, 5(1):5415, 2014.

\bibitem{Goh:2007aa}
Kwang-Il Goh, Michael~E. Cusick, David Valle, Barton Childs, Marc Vidal, and
  Albert-L{\'a}szl{\'o} Barab{\'a}si.
\newblock The human disease network.
\newblock {\em Proceedings of the National Academy of Sciences}, 104(21):8685,
  05 2007.

\bibitem{art-gtbgak12}
Li~Gong, Caroline~F. Thorn, Monica~M. Bertagnolli, Tilo Grosser, Russ~B.
  Altman, and Teri~E. Klein.
\newblock Celecoxib pathways: Pharmacokinetics and pharmacodynamics.
\newblock {\em Pharmacogenetics and genomics}, 22(4):310--318, 2012.

\bibitem{art-glxpztzczzjjwl20}
Luyao Guan, Zhang Li, Feifei Xie, Yuzhi Pang, Chenyun Zhang, Haosha Tang, Hao
  Zhang, Chun Chen, Yaying Zhan, Ting Zhao, Hongyuan Jiang, Xiaona Jia,
  Yuexiang Wang, and Yuan Lu.
\newblock Oncogenic and drug-sensitive ret mutations in human epithelial
  ovarian cancer.
\newblock {\em Journal of experimental and clinical cancer research}, 39(53),
  2020.

\bibitem{art-gzcyvwyl10}
Christopher~P. Gully, Fanmao Zhang, Jian Chen, James~A. Yeung, Guermarie
  Velazquez-Torres, Edward Wang, Sai-Ching~Jim Yeung, and Mong-Hong Lee.
\newblock Antineoplastic effects of an aurora b kinase inhibitor in breast
  cancer.
\newblock {\em Molecular cancer}, 22:9--42, 2010.

\bibitem{Guo:2017aa}
Wei-Feng Guo, Shao-Wu Zhang, Ze-Gang Wei, Tao Zeng, Fei Liu, Jingsong Zhang,
  Fang-Xiang Wu, and Luonan Chen.
\newblock Constrained target controllability of complex networks.
\newblock {\em Journal of Statistical Mechanics: Theory and Experiment},
  2017(6):063402, 2017.

\bibitem{art-hlwl18}
Liangbo Han, Huaqiang Liu, Jinfeng Wu, and Jinkai Liu.
\newblock mir-126 suppresses invasion and migration of malignant glioma by
  targeting mature t cell proliferation 1 (mtcp1).
\newblock {\em Medical science monitor}, 24:6630--6637, 2018.

\bibitem{art-hlthlkkmcgswsmdsgs06}
Liz~Y. Han, Charles~N. Landen, Jose~G. Trevino, Jyotsnabaran Halder, Yvonne~G.
  Lin, Aparna~A. Kamat, Tae-Jin Kim, William~M. Merritt, Robert~L. Coleman,
  David~M. Gershenson, William~C. Shakespeare, Yihan Wang, Raji Sundaramoorth,
  Chester~A. Metcalf~3rd, David~C. Dalgarno, Tomi~K. Sawyer, Gary~E. Gallick,
  and Anil~K. Sood.
\newblock Antiangiogenic and antitumor effects of src inhibition in ovarian
  carcinoma.
\newblock {\em Cancer research}, 66(17):8633--8639, 2006.

\bibitem{art-hlm08}
Aiwu~Ruth He, Andreas~Peter Lindenberg, and John~Lindsay Marshall.
\newblock Biologic therapies for advanced pancreatic cancer.
\newblock {\em Expert review of anticancer therapy}, 8(8):1331--1338, 2008.

\bibitem{art-halbssd20}
Kathrin Hedegger, Hana Alg\"{u}l, Marina Lesina, Andreas Blutke, Roland~M.
  Schmid, Marlon~R. Schneider, and Maik Dahlhoff.
\newblock Unraveling erbb network dynamics upon betacellulin signaling in
  pancreatic ductal adenocarcinoma in mice.
\newblock {\em Molecular oncology}, 2020.

\bibitem{art-hb06}
K.~Heyninck and R.~Beyaert.
\newblock A novel link between lck, bak expression and chemosensitivity.
\newblock {\em Oncogene}, 25(12):1693--1695, 2006.

\bibitem{art-hn08}
Stephen Hiscox and Robert~I. Nicholson.
\newblock Src inhibitors in breast cancer therapy.
\newblock {\em Expert opinion on therapeutic targets}, 12(6):757--767, 2008.

\bibitem{art-jfwkgdnf08}
Guangping Jiang, Tanya Freywald, Jarret Webster, Daniel Kozan, Ron Geyer, John
  DeCoteau, Aru Narendran, and Andrew Freywald.
\newblock In human leukemia cells ephrin-b-induced invasive activity is
  supported by lck and is associated with reassembling of lipid raft signaling
  complexes.
\newblock {\em Molecular cancer research}, 6(2):291--305, 2008.

\bibitem{art-jbwnrsc10}
N.~Johnson, J.~Bentley, L.-Z. Wang, D.~R. Newell, C.~N. Robson, G.~I. Shapiro,
  and N.~J. Curtin.
\newblock Pre-clinical evaluation of cyclin-dependent kinase 2 and 1 inhibition
  in anti-estrogen-sensitive and resistant breast cancer cells.
\newblock {\em British journal of cancer}, 102(2):342--350, 2010.

\bibitem{Kalman1963}
R.~E. Kalman, Y.~C. Ho, and K.~S. Narendra.
\newblock Controllability of linear dynamical systems.
\newblock {\em Contributions to Differential Equations}, 1:189--213, 1963.

\bibitem{art-kcgp17}
Krishna Kanhaiya, Eugen Czeizler, Cristian Gratie, and Ion Petre.
\newblock Controlling directed protein interaction networks in cancer.
\newblock {\em Scientific Reports}, 7, 2017.

\bibitem{Kitano:2002aa}
Hiroaki Kitano.
\newblock Computational systems biology.
\newblock {\em Nature}, 420(6912):206--210, 2002.

\bibitem{art-kbskkm11}
Judice Koh, Kevin Brown, Azin Sayad, Dahlia Kasimer, Troy Ketela, and Jason
  Moffat.
\newblock Colt-cancer: functional genetic screening resource for essential
  genes in human cancer cell lines.
\newblock {\em Nucleic acids research}, 40(D1):D957--D963, 2011.

\bibitem{art-lwjtf10}
Elaine~L. Leung, Janica~C. Wong, Mary~G. Johlfs, Benjamin~K. Tsang, and
  Ronald~R. Fiscus.
\newblock Protein kinase g type ialpha activity in human ovarian cancer cells
  significantly contributes to enhanced src activation and dna synthesis/cell
  proliferation.
\newblock {\em Molecular cancer research}, 8(4):578--591, 2010.

\bibitem{art-lgzlsyy09}
Hong~Zhao Li, Yan Gao, Xiu~Lan Zhao, Yi~Xin Liu, Bao~Cun Sun, Jie Yang, and Zhi
  Yao.
\newblock Effects of raf kinase inhibitor protein expression on metastasis and
  progression of human breast cancer.
\newblock {\em Molecular cancer research}, 7:832--840, 2009.

\bibitem{Liu:2019aa}
Xiangrong Liu, Zengyan Hong, Juan Liu, Yuan Lin, Alfonso
  Rodr{\'\i}guez-Pat{\'o}n, Quan Zou, and Xiangxiang Zeng.
\newblock Computational methods for identifying the critical nodes in
  biological networks.
\newblock {\em Briefings in Bioinformatics}, 21(2):486--497, 6/26/2020 2019.

\bibitem{Liu:2015aa}
Xueming Liu and Linqiang Pan.
\newblock Identifying driver nodes in the human signaling network using
  structural controllability analysis.
\newblock {\em IEEE/ACM Trans Comput Biol Bioinform}, 12(2):467--472, Mar-Apr
  2015.

\bibitem{art-lsb11}
Yang-Yu Liu, Jean-Jacques Slotine, and Albert-L{\'a}szl{\'o} Barab{\'a}si.
\newblock Controllability of complex networks.
\newblock {\em Nature}, 2011.

\bibitem{art-lfcyqzclxyzz20}
Yueze Liu, Mengyu Feng, Hao Chen, Gang Yang, Jiangdong Qiu, Fangyu Zhao, Zhe
  Cao, Wenhao Luo, Jianchun Xiao, Lei You, Lianfang Zheng, and Taiping Zhang.
\newblock Mechanistic target of rapamycin in the tumor microenvironment and its
  potential as a therapeutic target for pancreatic cancer.
\newblock {\em Cancer letters}, 485:1--13, 2020.

\bibitem{tri-mgh19}
{Massachusetts General Hospital}.
\newblock Cabozantinib in women with metastatic hormone-receptor-positive
  breast cancer, 2019.
\newblock Available at \url{https://clinicaltrials.gov/ct2/show/NCT01441947}.

\bibitem{sof-mn19}
MathNET.
\newblock Math.net numerics, 2019.
\newblock Available at \url{https://numerics.mathdotnet.com/}.

\bibitem{tri-mdacc05}
{M.D. Anderson Cancer Center}.
\newblock Gemcitabine and celecoxib in treating patients with metastatic
  pancreatic cancer, 2005.
\newblock Available at \url{https://clinicaltrials.gov/ct2/show/NCT00068432}.

\bibitem{art-mtldclrkjdg06}
Rajshree~R. Mewani, Song Tian, Bihua Li, Malika~T. Danner, Theresa~D. Carr,
  Sung Lee, Aquilur Rahman, Usha~N. Kasid, Mira Jung, Anatoly Dritschilo, and
  Prafulla~C. Gokhale.
\newblock Gene expression profile by inhibiting raf-1 protein kinase in breast
  cancer cells.
\newblock {\em International journal of molecular medicine}, 17(3):457--463,
  2006.

\bibitem{art-miklsasa04}
Ron Milo, Shalev Itzkovitz, Nadav Kashtan, Reuven Levitt, Shai Shen-Orr, Inbal
  Ayzenshtat, Michal Sheffer, and Uri Alon.
\newblock Superfamilies of evolved and designed networks.
\newblock {\em Science}, 303(5663):1538--1542, 2004.

\bibitem{art-msikca02}
Ron Milo, Shai Shen-Orr, Shalev Itzkovitz, Nadav Kashtan, Dmitri Chklovskii,
  and Uri Alon.
\newblock Network motifs: Simple building blocks of complex networks.
\newblock {\em Science}, 298(5594):824--827, 2002.

\bibitem{Misselbeck:2019aa}
Karla Misselbeck, Silvia Parolo, Francesca Lorenzini, Valeria Savoca, Lorena
  Leonardelli, Pranami Bora, Melissa~J. Morine, Maria~Caterina Mione, Enrico
  Domenici, and Corrado Priami.
\newblock A network-based approach to identify deregulated pathways and drug
  effects in metabolic syndrome.
\newblock {\em Nature Communications}, 10(1):5215, 2019.

\bibitem{Molnar:2013aa}
F.~Moln{\'a}r, S.~Sreenivasan, B.~K. Szymanski, and G.~Korniss.
\newblock Minimum dominating sets in scale-free network ensembles.
\newblock {\em Scientific Reports}, 3(1):1736, 2013.

\bibitem{art-mnh08}
Liam Morgan, Robert~I. Nicholson, and Stephen Hiscox.
\newblock Src as a therapeutic target in breast cancer.
\newblock {\em Endocrine, metabolic and immune disorders drug targets},
  8(4):273--278, 2008.

\bibitem{art-m02}
Charles Morris.
\newblock The role of egfr-directed therapy in the treatment of breast cancer.
\newblock {\em Breast cancer research and treatment}, 75:S51--S59, 2002.

\bibitem{tri-nci06}
{National Cancer Institute (NCI)}.
\newblock Sorafenib in treating patients with metastatic breast cancer, 2006.
\newblock Available at \url{https://clinicaltrials.gov/ct2/show/NCT00096434}.

\bibitem{art-npvkmj01}
J.~Novotn{\'y}, L.~Petruzelka, J.~Vedralov{\'a}, Z.~Kleibl, B.~Matous, and
  L.~Juda.
\newblock Prognostic significance of c-erbb-2 gene expression in pancreatic
  cancer patients.
\newblock {\em Neoplasma}, 48(3):188--191, 2001.

\bibitem{Ochsner:2019aa}
Scott~A. Ochsner, David Abraham, Kirt Martin, Wei Ding, Apollo McOwiti, Wasula
  Kankanamge, Zichen Wang, Kaitlyn Andreano, Ross~A. Hamilton, Yue Chen,
  Angelica Hamilton, Marin~L. Gantner, Michael Dehart, Shijing Qu, Susan~G.
  Hilsenbeck, Lauren~B. Becnel, Dave Bridges, Avi Ma'ayan, Janice~M. Huss,
  Fabio Stossi, Charles~E. Foulds, Anastasia Kralli, Donald~P. McDonnell, and
  Neil~J. McKenna.
\newblock The signaling pathways project, an integrated `omics knowledgebase
  for mammalian cellular signaling pathways.
\newblock {\em Scientific Data}, 6(1):252, 2019.

\bibitem{art-pm09}
Sumanta~Kumar Pal and Joanne Mortimer.
\newblock Triple-negative breast cancer: Novel therapies and new directions.
\newblock {\em Mauritas}, 63(4):269--274, 2009.

\bibitem{sof-p19}
Victor Popescu.
\newblock Geneticalgnetcontrol, 2020.
\newblock Available at \url{https://github.com/vicbgdn/GeneticAlgNetControl},
  version 1.0.

\bibitem{Rancati:2018aa}
Giulia Rancati, Jason Moffat, Athanasios Typas, and Norman Pavelka.
\newblock Emerging and evolving concepts in gene essentiality.
\newblock {\em Nature Reviews Genetics}, 19(1):34--49, 2018.

\bibitem{tri-rp10}
{Rigel Pharmaceuticals}.
\newblock Efficacy and safety study of fostamatinib tablets to treat b-cell
  lymphoma, 2010.
\newblock Available at \url{https://clinicaltrials.gov/ct2/show/NCT00446095}.

\bibitem{Saqi:2016aa}
Mansoor Saqi, Johann Pellet, Irina Roznovat, Alexander Mazein, St{\'e}phane
  Ballereau, Bertrand De~Meulder, and Charles Auffray.
\newblock Systems medicine: The future of medical genomics, healthcare, and
  wellness.
\newblock {\em Methods Mol Biol}, 1386:43--60, 2016.

\bibitem{tri-sahsmzu23}
{Second Affiliated Hospital, School of Medicine, Zhejiang University}.
\newblock Gemcitabine and celecoxib combination therapy in treating patients
  with r0 resection pancreatic cancer (gcrp), 2023.
\newblock Available at \url{https://clinicaltrials.gov/ct2/show/NCT03498326}.

\bibitem{art-slmqlaylc20}
Fugen Shangguan, Yan Liu, Li~Ma, Guiwu Qu, Qing Lv, Jing An, Shude Yang, Bin
  Lu, and Qizhi Cao.
\newblock Niclosamide inhibits ovarian carcinoma growth by interrupting
  cellular bioenergetics.
\newblock {\em Journal of Cancer}, 11(12):3454--3466, 2020.

\bibitem{art-stscdzwawexmgztmcvpeskcc20}
Sunita Shankar, Jean Ching-Yi Tien, Ronald~F. Siebenaler, Seema Chugh,
  Vijaya~L. Dommeti, Sylvia Zelenka-Wang, Xiao-Ming Wang, Ingrid~J. Apel,
  Jessica Waninger, Sanjana Eyunni, Alice Xu, Malay Mody, Andrew Goodrum,
  Yuping Zhang, John~J. Tesmer, Rahul Mannan, Xuhong Cao, Pankaj Vats,
  Sethuramasundaram Pitchiaya, Stephanie~J. Ellison, Jiaqi Shi, Chandan
  Kumar-Sinha, Howard~C. Crawford, and Arul~M. Chinnaiyan.
\newblock An essential role for argonaute 2 in egfr-kras signaling in
  pancreatic cancer development.
\newblock {\em Nature communications}, 11(1):2817, 2020.

\bibitem{art-sberch09}
Arun Sharma, Jennifer Belna, Joseph Espat, Gustavo Rodriguez, Vernon~T. Cannon,
  and Jean~A. Hurteau.
\newblock Effects of omega-3 fatty acids on components of the transforming
  growth factor beta-1 pathway: Implication for dietary modification and
  prevention in ovarian cancer.
\newblock {\em American journal of obstetrics and gynecology},
  200(5):516.e1--516.36, 2009.

\bibitem{tri-skcccjh20}
{Sidney Kimmel Comprehensive Cancer Center at Johns Hopkins}.
\newblock Clinical trial of combined fostamatinib and paclitaxel in ovarian
  cancer, 2020.
\newblock Available at \url{https://clinicaltrials.gov/ct2/show/NCT03246074}.

\bibitem{art-ssp08}
D.~Srinivasan, J.~T. Sims, and R.~Plattner.
\newblock Aggressive breast cancer cells are dependent on activated abl kinases
  for proliferation, anchorage-independent growth and survival.
\newblock {\em Oncogene}, 27(8):1095--1105, 2008.

\bibitem{art-samsubwbbabp05}
Jens Standop, Mahefatiana Andrianifahanana, Nicolas Moniaux, Matthias
  Schneider, Alexis Ulrich, Randall~E. Brand, James~L. Wisecarver, Julia~A.
  Bridge, Markus~W. B{\"u}chler, Thomas~E. Adrian, Surinder~K. Batra, and
  Parviz~M. Pour.
\newblock Erbb2 growth factor receptor, a marker for neuroendocrine cells?
\newblock {\em Pancreatology}, 5(1):44--58, 2005.

\bibitem{Tian:2012aa}
Q~Tian, N~D Price, and L~Hood.
\newblock Systems cancer medicine: towards realization of predictive,
  preventive, personalized and participatory (p4) medicine.
\newblock {\em J Intern Med}, 271(2):111--121, Feb 2012.

\bibitem{art-vrmbsslctlplpyhcwmcned20}
Halena~R. VanDeusen, Johnny~R. Ramroop, Katherine~L. Morel, Songyi Bae,
  Anjali~V. Sheahan, Zoi Sychev, Nathan~A. Lau, Larry~C. Cheng, Victor~M. Tan,
  Zhen Li, Ashley Petersen, John~K. Lee, Jung~Wook Park, Rendong Yang,
  Justin~H. Hwang, Ilsa Coleman, Owen~N. Witte, Colm Morrissey, Eva Corey,
  Peter~S. Nelson, Leigh Ellis, and Justin~M. Drake.
\newblock Targeting ret kinase in neuroendocrine prostate cancer.
\newblock {\em Molecular Cancer Research}, 2020.

\bibitem{pro-hss08}
G{\"a}el Varoquaux, Travis Vaught, and Jarrod Millman, editors.
\newblock {\em Exploring network structure, dynamics, and function using
  NetworkX}, Proceedings of the 7th Python in Science Conference (SciPy2008),
  2008.

\bibitem{Whitley:2012aa}
Darrell Whitley and Andrew~M. Sutton.
\newblock Genetic algorithms --- a survey of models and methods.
\newblock In Grzegorz Rozenberg, Thomas B\"{a}ck, and Joost~N. Kok, editors,
  {\em Handbook of Natural Computing}, pages 637--671. Springer Berlin
  Heidelberg, Berlin, Heidelberg, 2012.

\bibitem{art-wweptdbmg00}
Jon~R. Wiener, T.~Christopher Windham, Veronica~C. Estrella, Nila~U. Parikh,
  Peter~F. Thall, Michael~T. Deavers, Robert~C. Bast, Gordon~B. Mills, and
  Gary~E. Gallick.
\newblock Activated src protein tyrosine kinase is overexpressed in late-stage
  human ovarian cancers.
\newblock {\em Gynecologic oncology}, 88(1):73--79, 2003.

\bibitem{art-wfglmgsjlsanilmgwcclpkw18}
David~S. Wishart, Yannick~D. Feunang, An~C. Guo, Elvis~J. Lo, Ana Marcu,
  Jason~R. Grant, Tanvir Sajed, Daniel Johnson, Carin Li, Zinat Sayeeda,
  Nazanin Assempour, Ithayavani Iynkkaran, Yifeng Liu, Adam Maciejweski, Nicola
  Gale, Alex Wilson, Lucy Chin, Ryan Cummings, Diana Le, Allison Pon, Craig
  Knox, and Michael Wilson.
\newblock Drugbank 5.0: A major update to the drugbank database for 2018.
\newblock {\em Nucleic Acids Research}, 46:D1074--D1082, 2018.

\bibitem{art-xnsrnakb09}
Zangwei Xu, Kumiko Nagashima, Dongyu Sun, Thomas Rush, Alan Northrup, Jannik~N.
  Andersen, Ilona Kariv, and Ekaterina~V. Bobkova.
\newblock Development of high-throughput tr-fret and alphascreen assays for
  identification of potent inhibitors of pdk1.
\newblock {\em Journal of biomolecular screening}, 14(10):1257--1262, 2009.

\bibitem{art-ywt20}
T.-T. Yu, C.-Y. Wang, and R.~Tong.
\newblock Erbb2 gene expression silencing involved in ovarian cancer cell
  migration and invasion through mediating mapk1/mapk3 signaling pathway.
\newblock {\em European review for medical and pharmacological sciences},
  24(10):5267--5280, 2020.

\bibitem{art-zdnlpm03}
Huiyan Zeng, Kaustubh Datta, Matthias Neid, Jinping Li, Sareh Parangi, and
  Debabrata Mukhopadhyay.
\newblock Requirement of different signaling pathways mediated by insulin-like
  growth factor-i receptor for proliferation, invasion, and vpf/vegf expression
  in a pancreatic carcinoma cell line.
\newblock {\em Biochemical and biophysical research communications},
  302(1):46--55, 2003.

\bibitem{Zhang:2017aa}
Wei Zhang, Jeremy Chien, Jeongsik Yong, and Rui Kuang.
\newblock Network-based machine learning and graph theory algorithms for
  precision oncology.
\newblock {\em npj Precision Oncology}, 1(1):25, 2017.

\bibitem{art-zzjwz20}
H.-W. Zhao, N.~Zhou, F.~Jin, R.~Wang, and J.-Q. Zhao.
\newblock Metformin reduces pancreatic cancer cell proliferation and increases
  apoptosis through mtor signaling pathway and its dose-effect relationship.
\newblock {\em European review for medical and pharmacological sciences},
  24(10):5336--5344, 2020.

\bibitem{art-zwyzjxxyfwchzlywnvqd20}
Yan Zheng, Chao Wu, Jimeng Yang, Yue Zhao, Huliang Jia, Min Xue, Da~Xu, Feng
  Yang, Deliang Fu, Chaoqun Wang, Beiyuan Hu, Ze~Zhang, Tianen Li, Shican Yan,
  Xuan Wang, Peter~J. Nelson, Christiane Bruns, Lunxiu Qin, and Qiongzhu Dong.
\newblock Insulin-like growth factor 1-induced enolase 2 deacetylation by hdac3
  promotes metastasis of pancreatic cancer.
\newblock {\em Signal transduction and targeted therapy}, 5(1):53, 2020.

\bibitem{Zhou:2014aa}
XueZhong Zhou, J{\"o}rg Menche, Albert-L{\'a}szl{\'o} Barab{\'a}si, and Amitabh
  Sharma.
\newblock Human symptoms--disease network.
\newblock {\em Nature Communications}, 5(1):4212, 2014.

\end{thebibliography}

\end{document}